\documentclass[12pt]{article}
\usepackage{mathtools,float}
\usepackage{caption}
\usepackage{subcaption}
\allowdisplaybreaks[4]
\usepackage[dvipsnames]{xcolor}
\usepackage[T1]{fontenc}
\usepackage[margin=1in]{geometry}
\usepackage{amsmath,amsthm,amssymb}
\usepackage[authoryear]{natbib}
\usepackage{booktabs}
\usepackage{graphicx}
\usepackage{amsthm}
\newtheorem{assumption}{Assumption}
\newtheorem{proposition}{Proposition}
\newtheorem{remark}{Remark}

\newtheorem{rmk_app}{Remark}[section]
\usepackage[backref=page]{hyperref}
\definecolor{darkblue}{rgb}{0, 0, 0.5}
\hypersetup{
     colorlinks = true,
     citecolor = Maroon,
     urlcolor = darkblue,
     linkcolor = darkblue
}

\usepackage[normalem]{ulem}
\usepackage{threeparttable}
\usepackage{comment}
 \usepackage{soul}

\usepackage{comment}

\begin{document}

\title{Inference with few treated units\footnote{The authors are grateful to Avi Feller, Andreas Hagemann, Aureo de Paula, Chris Hansen, Chris Taber, Federico Bugni, James MacKinnon, Jon Roth, Marinho Bertanha, Matias Cattaneo, Matt Webb, Michael Leung, Tim Armstrong, Tim Conley, Vitor Possebom, Xavier D'Haultfoeuille, and participants of the Workshop on Quantitative Methods (University of Chicago), the econometrics seminar at Warwick University, and the Methods Workshop at the Kellogg Institute (University of Notre Dame) for their helpful comments and suggestions. Bruno Ferman gratefully acknowledges financial support from CNPq and FAPESP. }
}

\author{Luis Alvarez\thanks{University of São Paulo. Email: \url{luis.alvarez@usp.br}}\qquad Bruno Ferman\thanks{Sao Paulo School of Economics-FGV. Email: \url{bruno.ferman@fgv.br}} \qquad Kaspar W\"uthrich\thanks{University of Michigan; CESifo. Email: \url{kasparwu@umich.edu}}}

\date{First draft: April 26, 2025 \\ This draft: May 6, 2026 }

\maketitle

\begin{abstract} 
In many causal inference applications, only one or a few units (or clusters of units) are treated. An important challenge in such settings is that standard inference methods relying on asymptotic theory may be unreliable, even with large total sample sizes. This survey reviews and categorizes inference methods designed to accommodate few treated units, considering cross-sectional and panel data methods. We discuss trade-offs and connections between different approaches. In doing so, we propose slight modifications to improve the finite-sample performance of some methods, and we also provide theoretical justifications for existing heuristic approaches that have been proposed in the literature.

\end{abstract}


\section{Introduction}

In many causal inference applications, only one or a few units (or clusters of units) are treated. Examples include comparative case studies based on aggregate panel data, difference-in-differences (DiD) designs with few treated clusters, and randomized controlled trials (RCTs) with expensive treatments, among others. A key challenge in such settings is that standard inference methods based on asymptotic theory may be unreliable, even when the total number of units is large. For example, in simple treatment-control comparisons with few treated units (or clusters of units), the treated observations have a high leverage, which can lead to severe downward biases in conventional robust and cluster-robust standard errors \citep{Chesher,Carter,young2016improved,young_QJE,MacKinnonStata}.

This survey  reviews and categorizes the fast-growing literature proposing inference methods that are specifically designed to accommodate settings with few treated units. We consider both cross-sectional and panel data methods. We discuss the main assumptions that different inference methods rely on, explain the rationale behind them, and discuss their theoretical properties. We also emphasize connections and trade-offs between different approaches. 

The existing inference methods can be first categorized into two main groups based on the source of uncertainty they consider: model-based and design-based inference methods. Model-based methods focus on the uncertainty coming from sampling the potential outcomes from an infinite super-population. Design-based methods focus on the uncertainty coming from the randomness in the treatment assignment.\footnote{ Our distinction between model-based and design-based approaches is consistent with usage of these terms in early statistical literature \citep{Sarndal1978}, as well as more recent discussion in Econometrics \citep{baker2025}. The term ``sampling-based'' is also used to cover our notion of ``model-based'' in some settings \citep[e.g.][]{Abadie_finitepop}. We can also consider methods that account for uncertainty in \emph{both} potential outcomes and the treatment assignment \citep{Abadie_finitepop,Abadie2022}. Since this approach also requires knowledge on the treatment assignment mechanism, we discuss these methods along with design-based approaches in Section \ref{Sec: design_based}.}$^,$\footnote{We  only cover frequentist inference, as this is more common in econometrics. However, we should note that some recent papers have suggested Bayesian  approaches for inference in settings with few treated units \citep[e.g.,][]{Pang2022,benmichael2022,martinez2024}.}

The main focus of this review is on model-based inference methods, as these methods are more prevalent in econometrics. We organize the presentation of the different model-based inference methods according to data availability. We first consider methods that are valid even in the extreme case in which there is only one treated unit and one treated period. The main challenge in this setting is that there is very limited information on the distribution of potential outcomes of the treated unit in the treated period. Therefore, the solutions proposed in the literature rely on extrapolating \emph{cross-sectional} information from the untreated units or \emph{time-series} information from untreated periods to assess uncertainty about the distribution of the treated potential outcomes. The choice between these alternatives should depend on the data availability and on the assumptions one is willing to make regarding the distribution of the errors in modeling the potential outcomes. Methods that rely on cross-sectional information generally allow for unrestricted time-series dependence, but rely on assumptions that restrict how the distribution of errors varies across treated and control units and often require a large number of control units.  In contrast, methods that rely on time-series information allow for cross-sectional dependence and heteroskedasticity in the errors, but instead impose restrictions on the time-series properties of the errors (such as stationarity and weak dependence) and typically require a large number of untreated periods. Both types of methods typically also rely on strong restrictions on treatment effect heterogeneity.

We then consider alternatives that are valid when there are few (but more than one) treated units, when there are many treated periods, or when there are many individual-level observations within each cluster. We discuss how the additional available information in each setting can be used to relax some of the strong assumptions required for valid inference in the extreme case with one treated unit and period. Nevertheless, these methods generally require alternative assumptions in other dimensions, and still typically require stronger assumptions than standard methods that are valid with many treated and many control units. For example, with multiple treated units, there are methods that can relax assumptions limiting heteroskedasticity and treatment effect heterogeneity while instead relying on symmetry assumptions on the errors and treatment effects. Moreover, while these methods control size with a fixed number of treated units, they may have low power when the number of treated units is very small. Therefore, in a scenario where the assumptions underlying different approaches may seem reasonable, methods designed specifically for settings with a single treated unit and period might be preferred due to power considerations when the number of treated units is very small (but greater than one). 

Finally, we also discuss design-based approaches for inference. The theoretical justifications for design-based approaches are conceptually different from those for model-based approaches. Moreover, the focus is typically on different target parameters. This makes it difficult to directly compare them to model-based approaches. Nevertheless, we argue that design-based methods that are valid with few treated units share similar limitations as their model-based counterparts.

In categorizing different approaches for inference in settings with few treated units, we also show asymptotic equivalence between some of the methods considered in the literature. In some cases, this provides a theoretical justification for methods that have been only heuristically justified. For example, we show that, in some settings, a wild bootstrap with the null imposed is asymptotically equivalent to an approximate randomization test based on sign-changes when the number of control units goes to infinity (keeping the number of treated units fixed).\footnote{\cite{Canay_wild_bootstrap} derive conditions under which wild-cluster bootstrap is valid with a finite number of clusters. However, as we discuss in Section \ref{Sec: N1>1}, their results do not directly apply to  treatment-control comparisons or DiD applications in which we use the wild-cluster bootstrap at the unit level.} Moreover, we propose slight modifications to improve the finite-sample performance of some methods. The formal details are in the appendix.

This survey complements the excellent existing surveys and books on causal panel data methods where settings with few treated units are ubiquitous \citep[e.g.,][among others]{abadie2021using,dechaisemartin2022twoway,roth2023what,arkhangelsky2024causal,dechaisemartin2024book} and the surveys and guides to practice regarding inference in regression models with clusters \citep[e.g.,][among others]{cameron2015practitioner,mackinnon2023cluster}. Unlike  these references, we focus on the inferential challenges arising from the presence of few treated units (or clusters of units) across a wide variety of settings and methods. This allows us to highlight connections and common principles and, as a byproduct, yield new formal justifications for existing heuristic procedures and variants of methods with improved finite sample performance. Many of the ideas and methods we discuss are generic and can be used in conjunction with a wide variety of panel and cross-sectional causal inference methods. We organize the literature based on conceptual aspects, such as the type of uncertainty, and practical aspects, such as the number of periods and treated units, to make the survey easy to navigate and useful for practitioners across a variety of fields and theoretical researchers alike.

In Section \ref{Sec: example}, we start with a simple example that illustrates why standard methods that are asymptotically valid with many treated and many control units typically fail in settings with few treated units.  This example also clarifies that the challenges for inference we discuss arise when the \textit{number} of treated units is small and not necessarily when the \textit{share} of treated units is small.   Sections \ref{Sec: model based} and \ref{Sec: model-based inference methods} present the model-based setting and inference methods.  We then discuss in Section \ref{Sec: design_based} the use of design-based methods in settings with few treated units. Section \ref{recommendations} provides recommendations for applied researchers, while Section \ref{directions} concludes with directions for future research.

\section{Why standard methods fail with few treated units: A simple example} \label{Sec: example}

We consider a simple example that illustrates why standard inference methods typically fail in settings with few treated units and the main challenges for inference in such settings. As we will see below, the insights from this simple example generalize to many other causal inference problems, including the estimation of treatment effects using regression and matching methods, and panel data methods such as DiD, factor and interactive fixed effects models, and synthetic control.

We are interested in estimating the effect of a binary treatment $D_j$ on an outcome $Y_j$. We consider the regression model,
$$
Y_j=\mu+\tau D_j+\eta_j,\quad \mathbb{E}[\eta_j | D_j]=0,
$$
where $\tau$ is the treatment effect of interest.\footnote{We consider in detail a potential outcomes framework in Section \ref{Sec: model based}.} There are $N_1$ treated and $N_0$ control units. The outcomes of the treated (control) units are independently drawn from the distribution of $Y_j|D_j=1$ ($Y_j|D_j=0$).  To illustrate the issues with few treated units, we can consider an asymptotic regime where $N_1$ is fixed and $N_0\rightarrow \infty$. The OLS estimator of $\tau$ is equal to the difference between the treated and control mean,
\begin{eqnarray}
    \hat \tau =\frac{1}{N_1} \sum_{j \in \mathcal{J}_1}Y_j -  \frac{1}{N_0} \sum_{j \in \mathcal{J}_0} Y_j = \tau + \frac{1}{N_1} \sum_{j \in \mathcal{J}_1}\eta_j -  \frac{1}{N_0} \sum_{j \in \mathcal{J}_0} \eta_j,
\end{eqnarray}
where $\mathcal{J}_d\equiv \{j:D_j=d\}$ for $d=0,1$.
Defining $\sigma^2_d = \mathbb{V}(\eta_j|D_j=d)$ for $d=0,1$, the variance of $\hat\tau$ is 
\begin{eqnarray}
    \mathbb{V}(\hat \tau) =\frac{\sigma^2_1}{N_1} + \frac{\sigma^2_0}{N_0}.\label{eq:true_variance}
\end{eqnarray}
We allow for heteroskedasticity, that is, $\sigma^2_0$ can be different from $ \sigma^2_1$.  Equation \eqref{eq:true_variance} reflects the fact that we are estimating two means (for the treated and control group) to construct an estimator for  $\tau$, so $\mathbb{V}(\hat \tau)$ is the sum of the variances of  the treated and control averages.

The standard approach for making inferences on $\tau$ is to use a $t$-test based on the heteroskedasticity-robust variance estimator, 
\begin{eqnarray}
    \widehat{\mathbb{V}(\hat \tau)}_{\mbox{\tiny heterosk}} =\frac{\hat \sigma^2_1}{N_1} + \frac{\hat \sigma^2_0}{N_0},
\end{eqnarray}
where $\hat \sigma^2_a = \frac{1}{N_a} \sum_{j \in \mathcal{J}_a} \hat \eta^2_j$, for $a \in \{0,1\}$, and $\hat \eta_j$ is the OLS residual. A key feature of $\widehat{\mathbb{V}(\hat \tau)}_{\mbox{\tiny heterosk}}$ is that it only uses information from the treated units to estimate $\sigma^2_1$ and information from the control units to estimate $\sigma^2_0$. 

Under standard regularity conditions, as $\min\{N_1,N_0\} \rightarrow \infty$ (that is, we have many treated and many control observations), the OLS estimator $\hat \tau$ is asymptotically normal and
\begin{eqnarray}
    \frac{\hat \tau - \tau}{\sqrt{\widehat{\mathbb{V}(\hat \tau)}_{\mbox{\tiny heterosk}}}} \overset{d}{\rightarrow} N(0,1).
\end{eqnarray}

This result implies that we can conduct asymptotically valid inference for the null $H_0: \tau =0$ using the $t$-statistic $t_{\mbox{\tiny heterosk}}=\frac{\hat \tau}{\sqrt{\widehat{\mathbb{V}(\hat \tau)}_{\mbox{\tiny heterosk}}}}$ and standard normal critical values. We do not need to impose any assumptions on the relative magnitude of $\sigma_1^2$ and $\sigma_0^2$. Intuitively, this is because there is enough information to separately estimate $\sigma_1^2$ based on the treated units and $\sigma_0^2$ based on the control units when $N_1$ and $N_0$ are both large. 

Consider now a setting in which $\min\{N_1,N_0\} \not \rightarrow \infty$. As an extreme case, suppose that $N_1 = 1$ while $N_0 \rightarrow \infty$. The first thing to notice in this case is that, while $\hat \tau$ remains unbiased, it  will not be consistent and $\sqrt{N}$-asymptotically normal (even though $N=N_1+N_0 \rightarrow \infty$). In addition, the coefficient on the treatment dummy in this case would be such that the residual of the treated unit equals exactly zero. Therefore, $\hat \sigma^2_1=0$, implying $\widehat{\mathbb{V}(\hat \tau)}_{\mbox{\tiny heterosk}} = \frac{\hat \sigma^2_0}{N_0}$. Since the true variance of $\hat \tau$ is $\mathbb{V}(\hat \tau) =\sigma^2_1 + \frac{\sigma^2_0}{N_0}$, this means that $\widehat{\mathbb{V}(\hat \tau)}_{\mbox{\tiny heterosk}}$  underestimates $\mathbb{V}(\hat \tau)$. If $\sigma^2_0 = \sigma^2_1$, then the estimator for the variance of $\hat \tau$ is approximately $N$ times smaller than the true variance of $\hat \tau$. Intuitively, since there is only one treated observation, there is only enough information to estimate $\mathbb{E}[Y_j|D_j=1]$ (and, therefore, construct an estimator for $\tau$), but there is no information left to estimate $\sigma_1^2$ using only information from treated units.  

In settings with more than one (but few) treated units, this problem is less severe, but the standard normal approximation for $t_{\mbox{\tiny heterosk}}$ can still be very inaccurate. Table \ref{tab:simulations} provides an illustration based on a simple simulation study in which $Y_j \sim N(0,1)$ for all $j$. Note that this is a relatively favorable setting in that the estimator $\hat \tau$ is normally distributed even in finite samples. Moreover, we are in a setting in which errors are homoskedastic so that $\sigma_1^2=\sigma_0^2$.  We set $N_0=100$ and show results for $N_1\in \{1,\dots,5\}$. The nominal level is 5\%. While the problem is substantially more severe with $N_1 =1$, it is still relevant when $N_1>1$ but small. For example, with $N_1=5$, the true variance is still 32\% larger than the expected value of the heteroskedasticity-robust variance estimator, which leads to a rejection rate of 15\% instead of 5\%.  We stress that these simulations are only meant to illustrate the problem; there is no guarantee that the over-rejection is limited to $15\%$ when $N_1 = 5$ in other settings.

\begin{table}[]
    \centering
    \caption{Simulations with $Y_j \sim N(0,1)$} 
    \begin{tabular}{c|cc}
    $N_1$ & $\mathbb{V}(\hat \tau)/\mathbb{E}[\widehat{\mathbb{V}}(\hat \tau)]$ & Rejection rate \\ \hline
1 & 98.73 & 0.84 \\
2 & 2.86 & 0.35 \\
3 & 1.80 & 0.22 \\
4 & 1.41 & 0.16 \\
5 & 1.32 & 0.15 

    \end{tabular}
    
\begin{minipage}{0.8\textwidth}
\small \textit{Notes:} this table presents simulations results in which $N_0=100$, for different values of $N_1$. We set $Y_j \sim N(0,1)$ for all $j$. The first column presents the ratio between the variance of $\hat \tau$ and the expected value of the heteroskedasticity-robust estimator for this variance. The second column presents rejection rates for a 5\% nominal-level test using the $t$-statistic based on the heteroskedasticity-robust variance estimator, and the critical values  of a standard normal distribution. 
\end{minipage}
    \label{tab:simulations}
\end{table}

Interestingly, \textit{under this particular data generating process (DGP)} in which errors are normal and homoskedastic, using homoskedastic variance estimators leads to tests with approximately correct size, even when $N_1=1$. The homoskedastic variance estimator is 
\begin{eqnarray}
    \widehat{\mathbb{V}(\hat \tau)}_{\mbox{\tiny homosk}} =\hat \sigma^2 + \frac{\hat \sigma^2}{N_0},
\end{eqnarray}
where $\hat \sigma^2 = \frac{1}{N} \sum_{j=1}^N \hat \eta^2_j$. A notable difference relative to $\widehat{\mathbb{V}(\hat \tau)}_{\mbox{\tiny heterosk}}$ is that $\widehat{\mathbb{V}(\hat \tau)}_{\mbox{\tiny homosk}}$ uses information from both the treated and the control units to estimate the variance of the error (which here is assumed to be the same for treated and control units). Under those simplifying assumptions,  we have $\hat \tau \sim N\left( \tau,  \sigma^2 + \frac{ \sigma^2}{N_0} \right)$, so $\frac{\hat \tau - \tau}{\sqrt{ \widehat{\mathbb{V}(\hat \tau)}_{\mbox{\tiny homosk}}}} \overset{d}{\rightarrow} N(0,1)$ when $N_0 \rightarrow \infty$, even when $N_1=1$. Since $\sigma_0^2=\sigma_1^2$, we can consistently estimate $\sigma_1^2$ using the control units. In other words, we are \textit{extrapolating} information from the control units to learn something about the distribution of the errors of the treated units. We emphasize that   using $ \widehat{\mathbb{V}(\hat \tau)}_{\mbox{\tiny homosk}}$ does not yield valid inferences when the errors are heteroskedastic (even maintaining the normality assumption), because such extrapolation is not valid with heteroskedasticity. As we will see in Section \ref{Sec_N1_cross_section}, while the assumption that $\eta_j$ has the same distribution for treated and control observations plays an important role, normality assumptions are typically not required for valid inference when $N_1=1$. That said,  normality assumptions would be necessary for inference based on a $t$-test with homoskedastic standard errors.

\begin{remark}
The rationale from the example above is also valid when we consider DiD settings with few treated units, and we consider standard errors clustered at the unit level. In settings with no variation in treatment timing, computing the DiD estimator with cluster-robust standard errors at the unit level is (up to a degrees-of-freedom correction) numerically the same as computing a cross-section regression of $\bar Y_j^{\mbox{\tiny post}} - \bar Y_j^{\mbox{\tiny pre}}$ on a constant and  treatment dummy, with heteroskedasticity-robust standard errors, where  $\bar Y_j^{\mbox{\tiny post}}$ ($\bar Y_j^{\mbox{\tiny pre}}$) is the post (pre) treatment average for unit $j$.\footnote{See Equation 5 from \cite{ferman2019inference}.} In particular, for DiD settings with one treated cluster, we should expect standard errors clustered at the unit level to  massively underestimate the standard error of the estimator.\footnote{See, e.g., \cite{conley20211inference}, \cite{ferman2019inference}, \cite{MacKinnon2017,MacKinnon2018,MACKINNON2020435}  for theoretical justifications and simulations on the potential problems in using  cluster-robust standard errors in settings with few treated clusters, and \cite{Ferman_assessment} for examples of published papers that presented clustered standard errors in settings with one treated cluster. \cite{MacKinnon2017,MacKinnon2018,MACKINNON2020435} and \cite{ferman2019inference} also discuss potential pitfalls of wild-cluster bootstrap in settings with few treated clusters.}
Therefore, we recommend that applied researchers do not report cluster-robust standard errors in such settings. Instead, they should consider one of the methods discussed in Section \ref{Sec_N1} for inference.

\end{remark}

\begin{remark}
Standard  methods, such as inference based on heteroskedasticity-robust or cluster-robust variance estimators, are often asymptotically valid when  $\min \{N_1,N_0\} \rightarrow \infty$, even when treated units are a vanishing share of the total number of units (that is, $N_1/N \rightarrow 0$).\footnote{\cite{Janssen1997} shows this result for difference-in-means problems. Other examples in which we can construct $t$-tests that are asymptotically standard normal in settings with  $\min \{N_1,N_0\} \rightarrow \infty$ and $N_1/N \rightarrow 0$ include matching estimators \citep{AI_2006}, and the synthetic DiD estimator \citep{synthetic_did}.} Therefore, what matters to determine whether we are in a setting with few treated units is the \ul{number} of treated units, $N_1$, and not the \ul{proportion} of treated units, $N_1/N$.

\end{remark}

\section{Model-based setting} \label{Sec: model based}

\subsection{Notation and sources of uncertainty}

We consider settings in which we observe $N$ units, indexed by $j=1,\dots,N$, over $T$ periods, indexed by $t=1,\dots,T$. We discuss cross-sectional settings where $T=1$ and panel data settings where $T>1$. Our goal is to identify and estimate the causal effect of a binary treatment $D_{j,t}$ on an outcome of interest $Y_{j,t}$.
For ease of exposition, we focus on settings in which all treated units adopt the treatment in the same period and remain treated afterwards, so that we can define the treated and untreated potential outcomes as $Y_{j,t}(1)$ and $Y_{j,t}(0)$, respectively. However, the basic principles we discuss in the survey apply more generally. The observed outcome is related to the potential outcomes as
$Y_{j,t}=D_{j,t}Y_{j,t}(1)+(1-D_{j,t})Y_{j,t}(0)$. We may also observe a vector of covariates $X_{j,t}$. The causal effect of $D_{j,t}$ for unit $j$ in period $t$ is $\tau_{j,t}=Y_{j,t}(1)-Y_{j,t}(0)$, where $\tau_{j,t}$ could be random or fixed. In the following, we suppress the index $t$ whenever we describe cross-sectional methods. 

We first consider a model-based setting in which treatment assignment is fixed (or conditioned on), and the stochastic variation comes from the potential outcomes, $(Y_{j,t}(0),Y_{j,t}(1))$. Since treatment assignment is fixed and we are considering settings with no variation in treatment timing, we can define $\mathcal{J}_1$ ($\mathcal{J}_0$) as the set of treated (control) units, and $\mathcal{T}_1$ ($\mathcal{T}_0$) as the set of post- (pre-)treatment periods. We also let \(N_d = |\mathcal{J}_d|\) and \(T_d = |\mathcal{T}_d|\) for \(d \in \{0,1\}\), denote the number of control/treated observations and pre-/post-treatment periods, respectively.

This model-based perspective is appropriate in applications where there is a well-defined large population from which the sample is drawn.\footnote{See, e.g., \citet{roth2023what} for a discussion in the context of DiD methods.} The standard justification for this  interpretation is that there is a large population of units, and we are sampling a negligible fraction of it. Alternatively, if it makes sense to view the sample as the population of interest, we can interpret it as being drawn from a larger population -- the superpopulation -- representing different possible realizations of random variables that determine the outcomes of the units in the sample.  

This  alternative interpretation of sampling is not new: it dates back to at least since \cite{Haavelmo1944}, who wrote:
\begin{quote}
 There is no logical difficulty involved in considering the ``whole population as a sample,'' for the class of populations we are dealing with does \emph{not} consist of an infinity of different individuals, it consists of an infinity of possible \emph{decisions} which might be taken with respect to the
 value of $y$.
\end{quote}
and also appears in more recent discussions on the nature of uncertainty in structural analyses \citep[e.g.,][]{heckman2000causal,heckman2005scientific} and in Biostatistics \citep[e.g.,][Section 1.2]{hernan2010causal}.

This alternative interpretation of sampling is particularly relevant for justifying model-based approaches in applications with few treated units. For example, we often consider applications where a few states are treated, and we observe outcomes for all states. Moreover, we can even consider settings in which treatment would not be well-defined for the control units. For example, in the study of the economic impact of the German reunification by \citet{abadie2015comparative}, it is difficult to imagine other countries reunifying  \citep[see, e.g., the discussion in][]{abadie2021using}. Still, in such setting, we can consider uncertainty coming from the fact that, conditional on the German reunification occurring in 1989, there were infinite possible realizations of the random variables that determine the potential outcomes of the treated and control countries, from which we observe a single realization.

 Finally, model-based analyses can be interpreted as being conditional on the treatment assignment if the treatment assignment is stochastic,  allowing for explicitly incorporating treatment assignment mechanisms into the analysis \citep[see, e.g., the discussion in][]{Ferman_JASA}. Therefore, we can also use this setup to analyze settings in which treatment is also stochastic.

\subsection{Model, parameters of interest, and estimators} \label{Sec: model, parameters of interest, and estimators}
 
The literature typically focuses on treatment effects on the treated units in the treated period. A popular target parameter is the average treatment effect on the treated units (ATT), \begin{eqnarray} \label{eq_ATT}
    \tau^\ast \equiv \mathbb{E}\left[\frac{1}{N_1} \frac{1}{T_1} \sum_{j \in \mathcal{J}_1} \sum_{t \in \mathcal{T}_1} \tau_{j,t} \right],
\end{eqnarray}
where this expectation is taken over  the distribution of the potential outcomes in the population (or super-population). Most of our discussions remain relevant if we consider alternative target parameters, such as the sequence $\left\{\mathbb{E}\left[\frac{1}{N_1}\sum_{j \in \mathcal{J}_1}  \tau_{j,t} \right]  \right\}_{t \in \mathcal{T}_1}$.  We discuss other alternatives, such as inference on the \textit{realized} treatment effects, in Section \ref{Sec_Alternative_targets}.  

The key challenge when identifying and estimating parameters like the ATT is that the untreated potential outcomes are not observed for the treated units in the post-treatment period. By the definition of the potential outcomes, we can write the treatment effect $\tau_{j,t}$ for $j\in \mathcal{J}_1$ and $t\in \mathcal{T}_1$ as
$$
\tau_{j,t}=Y_{j,t}(1)-Y_{j,t}(0)=Y_{j,t}-Y_{j,t}(0).
$$
Thus, the key unknowns are the untreated potential outcomes for the treated units in the post-treatment period. 

Most of the model-based methods we analyze in this review postulate models for the potential outcomes that can be written as
\begin{eqnarray}
Y_{j,t}(0)=M_{j,t}+\epsilon_{j,t},\label{eq:basic_model}
\end{eqnarray}
where $M_{j,t}$ is a mean-predictor of $Y_{j,t}(0)$ and $\epsilon_{j,t}$ is an error term, which is typically assumed to have mean zero.\footnote{{If the model for the potential outcome is misspecified, $M_{j,t}$ can be interpreted as a pseudo-mean predictor. See further discussion in Appendix \ref{app_mis}.}} For example, in comparison-of-means applications (motivated by, for instance, the analysis of a randomized controlled trial in a model-based setting), we could consider $M_j = \mathbb{E}[Y_j(0)]$, where index $t$ is omitted since we are in a cross section. We present in Section 3.3 a series of examples that can be analyzed under this setup.

Given a model $M_{j,t}$, the estimator of the ATT will usually take the form 
\begin{eqnarray} \label{Eq: tau hat}
\hat \tau = \frac{1}{N_1} \frac{1}{T_1} \sum_{j \in \mathcal{J}_1} \sum_{t \in \mathcal{T}_1} Y_{j,t} - \hat M,
\end{eqnarray}
where $\hat M$ is an estimator for $\frac{1}{N_1} \frac{1}{T_1} \sum_{j \in \mathcal{J}_1} \sum_{t \in \mathcal{T}_1} M_{j,t}$. For methods in which we estimate treatment effects by imputing  $\hat M_{j,t}$ as the unobserved counterfactual for unit $j$    at period $t$, the estimator takes the form $\hat M = \frac{1}{N_1} \frac{1}{T_1} \sum_{j \in \mathcal{J}_1} \sum_{t \in \mathcal{T}_1} \hat M_{j,t}$. For the comparison-of-means example above, a natural estimator is $\hat M_j = \frac{1}{N_0} \sum_{j \in \mathcal{J}_0} Y_j$.

Given Equation \eqref{Eq: tau hat}, the estimation error $\hat{\tau}-\tau^\ast$ can be decomposed into three terms, 
\begin{eqnarray} \label{Eq_alpha_hat}
\hat \tau - \tau^\ast =     \left(\frac{1}{N_1} \frac{1}{T_1} \sum_{j \in \mathcal{J}_1} \sum_{t \in \mathcal{T}_1} \epsilon_{j,t} \right) + \left(\frac{1}{N_1} \frac{1}{T_1} \sum_{j \in \mathcal{J}_1} \sum_{t \in \mathcal{T}_1} M_{j,t} - \hat M \right) + \left(\frac{1}{N_1} \frac{1}{T_1} \sum_{j \in \mathcal{J}_1} \sum_{t \in \mathcal{T}_1} \tau_{j,t} - \tau^\ast \right).
\end{eqnarray}
The first term is the average of the errors in $Y_{j,t}(0)$ for the treated units in the post-treatment periods. The second term captures the estimation error of the counterfactual. Finally, the third term captures the heterogeneity in the treatment effects. Here we distinguish between two kinds of heterogeneity in the treatment effects, which are not mutually exclusive. 

\begin{enumerate}
    \item We call \textit{deterministic} heterogeneous treatment effects when $\mathbb{E}[\tau_{j,t}]$ varies with $j$ and/or $t$. This allows, for example, the possibility that treatment effect is expected to decrease or increase over time, or that certain treated units are expected to have larger effects relative to others. Importantly, note that this type of heterogeneous treatment effects does not contribute to the variance of $\hat \tau$. Still, the presence of deterministic heterogeneous treatment effects may be relevant for the validity of some inference methods. 

    \item We call \textit{stochastic} heterogeneous treatment effects when, for a given $j$ and $t$, $\tau_{j,t}$ is stochastic. This allows, for example, for settings in which treatment has an effect on the risk of an economic activity, so we  should expect $\mathbb{V}(Y_{j,t}(1)) \neq \mathbb{V}(Y_{j,t}(0))$, which is not possible if we only have deterministic heterogeneous treatment effects. We can also consider situations in which treatment effects may be lower or higher depending on a weather shock, where $\tau^\ast$ would be the average treatment effect once we integrate over these shocks. In those cases, this type of treatment effect heterogeneity contributes to the variance of $\hat \tau$ as an estimator of $\tau^\ast$.    
\end{enumerate}

Note that the expected value of the third term is zero by definition. We focus on cases in which the expected values of the first and second terms are also zero (or at least asymptote to zero when $N_0$ and/or $T_0$ increases). Under these assumptions, the estimator is asymptotically unbiased.  The main challenge for inference in settings with few treated units is that a Central Limit Theorem (CLT) cannot usually be used to reliably approximate the distributions of the first and third terms.\footnote{We discuss in Section \ref{Sec: exploiting CLT} and in Appendix \ref{sec: subsampling} two exceptions in which the distribution of $\hat \tau$ can be approximated with a CLT even in settings with few treated units. } Moreover, it is not possible to consistently estimate the variances of $\epsilon_{j,t}$ and $\tau_{j,t}$ using only information from the treated units. As a result, standard inference methods that rely on asymptotic normality of the estimator and on  normal approximations of studentized test statistics are not reliable with few treated units.

\begin{remark}

\label{Remark_naive}

We consider in Appendix \ref{app_disag} the case in which treatment is assigned at a unit level, but we have information on individual-level observations within units. We discuss there the idea of considering inference conditional on aggregate shocks.

\end{remark}

\subsection{Examples}
\label{examples_M}
Our framework encompasses popular methods in  cross-sectional analyses. For example:

\medskip

\noindent \textbf{Comparison of means/RCT.} In this example, $M_{j}$ is equal to the mean of the untreated potential outcome, $M_{j}=\mathbb{E}[Y_{j}(0)]$. $M_{j}$ can be estimated as $\hat M_j = \frac{1}{N_0} \sum_{j \in \mathcal{J}_0} Y_j$.

\medskip

\noindent \textbf{Regression.} Let $M_{j}=\mu_0+X_{j}'\beta_0$, for some observed variables $X_j$. In this case,  $\beta_{0}$ is the population regression parameter from a regression of $Y_{j}$ on $X_j$ in the subpopulation with $D_j=0$. $M_{j}$ can be estimated using the sample analog of this regression as $\hat{M_{j}}=\hat\mu_0+X_{j}'\hat\beta_0$, where $(\hat{\mu}_0,\hat\beta_0)$ are obtained from a regression of $Y_{j}$ on $X_j$ in the subpopulation with $D_j=0$.\footnote{ Another way to implement an estimator for the ATT in this case would be to run a linear regression of $Y_j$ on $D_j$, $X_j$, and interactions of $D_j$ and $X_j - \bar X$ \citep{ImbensWoodlrdige2009}. If we run a linear regression of $Y_j$ on $D_j$ and $X_j$, then the estimator would not necessarily have the format from Equation \ref{Eq: tau hat}. Still, the main takeaways regarding inference with few treated units would remain valid in this case (see Section \ref{remark_cov_ct} of Appendix \ref{app_exact} for an example). }

\medskip

\noindent \textbf{Matching.} Let $M_j= \mu_0(X_j)$, where $\mu_0(X_j)=\mathbb{E}[Y_{j}(0)|X_j]$. In this case, the $K-$nearest neighbor matching estimator would take the form $\hat M_j = \frac{1}{K} \sum_{q \in \mathcal{V}_j} Y_q$, where $\mathcal{V}_j$ is the set with the $K$ control units with values of $X_q$ closest to $X_j$. Other types of matching estimators could also be used \citep[see, e.g.,][]{Imbens_matching}.

\medskip

The framework also encompasses many popular panel data methods. For example:
\medskip

\noindent \textbf{Difference-in-differences.} DiD methods are typically motivated by a two-way fixed effects model for the untreated potential outcome, $M_{j,t} = \lambda_t + \mu_j$, where $\lambda_t$ is a time fixed effect and $\mu_j$ is an unit fixed effect. If all treated units start treatment at the same period, a canonical DiD estimator is $\hat \tau = \frac{1}{N_1} \sum_{j \in \mathcal{J}_1} \left( \bar Y_{j,post} - \bar Y_{j,pre} \right) - \frac{1}{N_0} \sum_{j \in \mathcal{J}_0} \left( \bar Y_{j,post} - \bar Y_{j,pre} \right)$, where $\bar Y_{j,post} = \frac{1}{T_1} \sum_{t \in \mathcal{T}_1}Y_{j,t}$ and $\bar Y_{j,pre} = \frac{1}{T_0} \sum_{t \in \mathcal{T}_0}Y_{j,t}$. In this case, we have that $$\hat M_{j,t}=\bar Y_{j,pre} + \frac{1}{N_0} \sum_{j' \in \mathcal{J}_0} \left(  Y_{j',t} - \bar Y_{j',pre} \right).$$ See \cite{dechaisemartin2022twoway} and \cite{roth2023what} for alternative DiD estimators and settings with more complicated adoption patterns.

\medskip

\noindent \textbf{Factor and interactive fixed effects approaches.} The interactive fixed effects model is $M_{j,t} = X_{j,t}\beta+ \lambda_t '\mu_j$, where $\lambda_t$ is a vector of time-varying factors and $\mu_j$ is a vector of unit-specific loadings. It nests the standard factor model $M_{j,t} = \lambda_t '\mu_j$ as a special case. A natural estimator of $M_{j,t}$ is $\hat{M}_{j,t}=X_{j,t}\hat\beta+ \hat\lambda_t '\hat\mu_j$, where $(\hat\beta, \hat\lambda_t, \hat\mu_j)$ are estimated based on untreated periods and untreated groups \citep[e.g.,][]{gobillon2016regional,xu2017generalized}. Alternatively, models with factor structures can be estimated using matrix completion techniques \citep[e.g.,][]{amjad2018robust,athey2021matrix}.  

\medskip

\noindent \textbf{Synthetic control.} This method has been considered under different assumptions on $M_{j,t}$, such as linear factor models, low-rank matrices, or autoregresive models. Synthetic control estimators of $M_{j,t}$ typically take the form $\hat{M}_{j,t}=\sum_{j\in \mathcal{J}_0}\hat{w}_jY_{j,t}$, where the weights $\{\hat{w}_j\}_{j\in \mathcal{J}_0}$ are obtained based on the pre-treatment period \citep[see, e.g.,][for a review]{abadie2021using}.

\section{Model-based inference methods}

\label{Sec: model-based inference methods}

\subsection{Methods that are valid even with $N_1=1$ and $T_1=1$} 

\label{Sec_N1}

We consider first inference methods that are valid even for an extreme case in which  $N_1=1$ and $T_1=1$. Therefore, in this section, we let $j=1$ be the treated unit, and $t=T$ be the post-treatment period. Importantly, though, these methods are usually also valid for settings with $N_1>1$ and/or $T_1 > 1$.

A distinctive feature of this extreme setting with $N_1=T_1=1$ is that there is only a single observation $Y_{j,t}$ that is treated. Therefore, we have one observation to estimate the treatment effect $\tau^\ast$, but not enough variation to quantify the uncertainty about the distribution of $Y_{1,T}(1)$ (which, following the decomposition in Equation \eqref{Eq_alpha_hat}, encompasses the 
uncertainty on  $\epsilon_{1,T}$ and $\tau_{1,T}$). The solutions that are valid even in such extreme settings then attempt to use information from the control units and/or the pre-treatment periods to learn about the  distribution of $Y_{1,T}(1)$. Therefore, in all cases, we need to impose assumptions that are stronger than those required by standard inference methods for settings with many treated and many controls units. For example, since the untreated units do not provide any information on the distribution of treatment effects, restrictions on  treatment effect heterogeneity are typically unavoidable.

In Sections \ref{Sec_N1_cross_section} and \ref{Sec_N1_time_series}, we consider settings with no stochastic treatment effect heterogeneity in which $\tau_{1,T}$ is treated as a fixed parameter. Therefore, in these cases we have that the ATT is given by $\tau^\ast = \tau_{1,T}$. In Section \ref{Sec_N1_cross_section}, we focus on methods that exploit the cross-section variation for inference, while in Section \ref{Sec_N1_time_series} we focus on methods that exploit time series variation. Then, in Section \ref{Sec_Alternative_targets}, we discuss alternative interpretations and possible ways to relax the homogeneous treatment effects assumption.

\subsubsection{Methods that exploit cross-sectional variation} \label{Sec_N1_cross_section}

\cite{conley20211inference} provide a leading example of a model-based method that exploits variation in the cross-section. They propose an inference method for DiD settings that is valid when $N_1$ is fixed (including the extreme case where $N_1=1$) and $N_0 \rightarrow \infty$. Their main method is valid when (i) $\{\epsilon_{j,t}\}_{t \in \mathcal{T}}$ is iid across $j$ and  (ii)  the treatment effects are homogeneous.\footnote{\cite{conley20211inference} also consider alternatives that relax both of these assumptions.} We discuss the interpretation of their method with stochastic heterogeneous treatment effects in Section \ref{Sec_Alternative_targets}. 

With $N_1=T_1=1$, the ATT from Equation \eqref{eq_ATT} is the (homogeneous) treatment effect for the treated unit in the post-treatment period, $\tau^\ast = \tau_{1,T}$, and the DiD estimator is given by $\hat \tau = \left[ Y_{1,T} - \bar Y_{1,pre}\right] - \frac{1}{N_0} \sum_{j \in \mathcal{J}_0}\left[Y_{j,T} - \bar Y_{j,pre} \right]$. Under standard regularity conditions, as $N_0 \rightarrow \infty$, we have that   $\hat \tau  = \tau^\ast + W_1 + \frac{1}{N_0} \sum_{j \in \mathcal{J}_0} W_j \overset{p}{\rightarrow} \tau^\ast + W_1$ as $N_0\rightarrow \infty$, where $W_j \equiv \epsilon_{j,T} - \frac{1}{T_0} \sum_{t \in \mathcal{T}_0} \epsilon_{j,t}$ is a linear combination of the errors. Therefore, under standard DiD identification assumptions ($\mathbb{E}[W_j] = 0$ for all $j$), the DiD estimator in this setting with one treated unit would be unbiased, but it would not be consistent. Moreover, the asymptotic distribution of $\hat \tau$ when $N_0 \rightarrow \infty$ will depend only on the errors of the treated unit, $W_1$.

The main intuition underlying \cite{conley20211inference}'s method is that the residuals of the controls $\{\widehat W_j \}_{j \in \mathcal{J}_0}$ asymptotically recover the distribution of $W_j$ in the control group. Then, under the assumption that $W_1$ has the same distribution as $W_j$ for $j \in \mathcal{J}_0$, this implies that we also recover the distribution of $W_1$. Therefore, we can construct confidence  intervals for $\tau^\ast$ that are asymptotically valid when $N_0 \rightarrow \infty$ using the quantiles of the distribution of $\{\widehat W_j \}_{j \in \mathcal{J}_0}$,  $CI = [\hat \tau - \hat Q_W(1-\gamma/2), \hat \tau - \hat Q_W(\gamma/2)$], where $\hat Q_W(u)$ is the $u$ empirical quantile of $\{\widehat W_j \}_{j \in \mathcal{J}_0}$. Likewise, we can construct a p-value 
\begin{eqnarray}
    p = \frac{1}{N_0}\sum_{j \in \mathcal{J}_0}\mathbf{1}\left\{|\widehat W_j|  \geq |\hat \tau - c|\right\}\,,
\end{eqnarray}
which is asymptotically valid when $N_0 \rightarrow \infty$ for a two-sided test that  $H_0: \tau^\ast = c$.

Crucially, since we rely on information on the control units to learn about the distribution of the errors of the treated unit, we need to rely on assumptions that link these two distributions. \cite{conley20211inference} in their standard implementation assume that $W_1$ and $W_j$ for $j \in \mathcal{J}_0$ have the same distribution. Moreover, we need a large number of control units  to consistently estimate the distribution of $W_j$ for $j \in \mathcal{J}_0$. This is achieved in their setting under the assumption that $W_j$ is iid across $j$, and considering an asymptotic approximation with $N_0 \rightarrow \infty$. 

Interestingly, by exploiting cross-section variation for inference, \cite{conley20211inference} do not have to restrict the time-series properties of $\epsilon_{j,t}$. The main idea is that, to derive the distribution of  $\hat \tau$ in this setting, we can essentially collapse the data into differences between post- and pre-treatment periods. Then this post- and pre-differences in the errors, $W_j$,  already incorporates any serial correlation in $\epsilon_{j,t}$ that is relevant to derive the distribution of $\hat \tau$. Moreover, since this collapsed data is essentially a comparison of means, this approach can also be used as a model-based inference approach in RCTs with few treated units.

The method proposed by \cite{conley20211inference} can also be used in settings with $N_1>1$ and/or $T_1>1$. In this case, we still cannot allow for stochastic treatment effects heterogeneity, but we can allow for deterministic treatment effects heterogeneity. Therefore, we continue to treat $\tau_{j,t}$  as deterministic parameters, but we allow them to vary with both $j$ and $t$, so that $\tau^\ast = \frac{1}{N_1} \frac{1}{T_1} \sum_{j \in \mathcal{J}_1}\sum_{t \in \mathcal{T}_1} \tau_{j, t}$ (see Section \ref{Sec: model, parameters of interest, and estimators} for more details on the distinction between these two types of treatment effects heterogeneity). With $N_1>1$ and/or $T_1>1$ we have that    $\hat \tau \overset{p}{\rightarrow} \tau^\ast + \frac{1}{N_1}\sum_{j \in \mathcal{J}_1} \dot W_j$ when $N_0 \rightarrow \infty$ and $N_1$ is fixed,
where, in this case, $\dot W_j  \equiv \frac{1}{T_1} \sum_{t \in \mathcal{T}_1} \epsilon_{j,t} - \frac{1}{T_0} \sum_{t \in \mathcal{T}_0} \epsilon_{j,t}$. Therefore, the asymptotic distribution of $\hat \tau$ depends only on the errors $\dot W_j$ of the treated, regardless of whether $\tau_{j,t}$ varies with $j$ or $t$. Since, under the assumptions considered above, we can recover the distribution of $\dot W_j$ for the treated units by extrapolating from information from the control units, the inference method is valid in this case, even with a small number of treated units.\footnote{With $N_1>1$, it is also possible to consider an alternative using a studentized test statistic, as proposed by \cite{MACKINNON2020435}. While with $N_1$ fixed it would not be possible to guarantee that the test is valid when there is treatment effect heterogeneity (even when $N_0 \rightarrow \infty$), this has important  advantages in settings with $N_1 \rightarrow \infty$, as discussed in Appendix \ref{app_perm_t} and Section \ref{Sec: dual}.}

As \cite{conley20211inference} noted, their inference procedure is asymptotically equivalent to a permutation test. We show in Appendix \ref{app_exact} that a slightly modified version of their method is exactly equivalent to a permutation test, being valid   with fixed $N_1$ and $N_0$ provided that $\{\epsilon_{j,t}\}_{t=1}^T$ is iid across $j$. The only additional assumption for validity with fixed $N_0$ is that we cannot have deterministic treatment effect  heterogeneity in the cross-section when we have more than one treated unit (which is allowed asymptotically when $N_0 \rightarrow \infty$). We present details on that in  Appendix \ref{app_exact}. The proposed alternative implementation of \cite{conley20211inference}'s method exhibits relevant advantages at essentially no cost. More specifically, it has the advantage of having finite-$N_0$ validity (under slightly stronger effect homogeneity assumptions), while being   asymptotically equivalent to the original implementation when $N_0 \rightarrow \infty$. Therefore, we recommend the use of this alternative implementation.  An important caveat of this approach when   both $N_0$ and $N_1$ are small is that the reference distribution for the test will have a small number of support points. This implies that, in a setting with $N_1=1$, the p-value would always be, by construction, greater than $\frac{1}{N_0 + 1}$. Therefore, at least $N_0 \geq 10$ ($N_0 \geq 20$) would be required to reject the null at a 10\% ($5\%$) significance level.

A number of papers propose similar approaches that exploit cross-sectional variation for inference with a single or few treated units under different sets of assumptions and for different settings. \cite{ferman2019inference} consider a DiD setting in which $W_j$ may exhibit heteroskedasticity based on observed variables. As a leading example, when $Y_{j,t}$ represents state $\times$ time aggregates, we should expect $W_j$ for states with larger populations to have relatively lower variances. As a result, the standard method of \cite{conley20211inference} over- (under-)rejects when the treated state is relatively small (large).\footnote{\label{Footnote: unconditional CT} If we assume a setting in which treatment assignment is also stochastic, and all units have uniform probability of being the treated one, then the method proposed by \cite{conley20211inference} would remain valid when we consider inference  \textit{unconditional} with respect to the treatment assignment. However, we would have size distortions for inference conditional on the populations of the treated states. See, e.g., \cite{ferman2019inference} for some arguments in favor of considering conditional inference in these settings.} 
To address this issue, \cite{ferman2019inference} propose to estimate this heteroskedasticity using control residuals. Then we can re-scale $\{\widehat W_j \}_{j \in \mathcal{J}_0}$ using the estimated heteroskedasticity and recover the distribution of $W_1$. This approach is asymptotically valid in  settings with $N_0 \rightarrow \infty$ and $N_1$ fixed (even when $N_1=1$) if we assume a scale-change model for the distribution of $W_j = \xi_j \times \sigma(X_j)$, where $\xi_j$ is iid across treated and control units, but $\mathbb{V}(W_j)$ depends on a set of observed covariates, $X_j$. In particular, in the example in which $Y_{j,t}$ represents state $\times$ time aggregates, this approach corrects for the fact that the errors of the treated state should have a higher variance than the errors of the control states, if the treated state is relatively smaller. Importantly, this approach still does not allow for {stochastic} heterogeneous treatment effects, and does not allow for heteroskedasticity in $W_j$ beyond the one that is estimable with the observed data.  We also note that the method proposed by \cite{ferman2019inference} relies crucially on $N_0 \rightarrow \infty$, which allows for consistent estimation of the heteroskedasticity function using the control units. As a result, the modifications in Appendix \ref{app_exact} that ensure the finite $N_0$ validity of \cite{conley20211inference}'s method do not guarantee the finite $N_0$ validity of  \cite{ferman2019inference}'s method. 

Still considering DiD settings, \cite{alvarez2023extensions} consider settings with variation in treatment timing, a topic that has been extensively studied in the  recent DiD literature (see \cite{dechaisemartin2022twoway} and \cite{roth2023what} for surveys). They  show that some recently proposed alternatives may also lead to substantial over-rejection when there are few treated units and extend the methods proposed by \cite{conley20211inference} and \cite{ferman2019inference} to accommodate variation in treatment timing. They also  derive uniform confidence bands for dynamic DiD specifications that are valid with fixed $N_1$, including the case  $N_1=1$. In another paper, \cite{alvarez2023inference} relax the independence assumption across units. They  show that the methods proposed by \cite{conley20211inference} and \cite{ferman2019inference} remain valid  under weak dependence in the cross section when $N_1=1$, and propose alternatives when $N_1 > 1$.

The main idea underlying \cite{conley20211inference}'s method has also been used for inference in combination with other estimators than DiD. For example, \cite{synthetic_did} (Algorithm 4) build on this idea to develop an approach for making inferences based on their Synthetic DiD estimator in applications with $N_1=1$, and \cite{ferman_matching} develops a related approach for inference based on matching estimators with few treated and many control units. 
In regression discontinuity designs,  \cite{CanayKamat2017} consider a related approach for testing the null hypothesis of covariate balance in an asymptotic regime where the number of observations around the threshold used in the analysis is fixed and the total number of observations in the sample increases.

There are also other alternative methods that are valid with $N_1=1$, even when  $N_0$ is fixed. For example, \cite{Donald} consider DiD settings with fixed $N_1$ and $N_0$, and derive the exact distribution of the $t$-statistic under normality and homoskedasticity assumptions on the errors (in contrast to \cite{conley20211inference}, who do not require normality assumptions). Finally, \cite{hagemann2020inference} proposes another alternative for settings with a finite number of heterogeneous clusters, where $N_1=1$. Each cluster is assumed to be large, so that the average error for each cluster is approximately Gaussian, and heteroskedasticity is allowed by imposing upper bounds on the ratio between the variance of the treated unit relative to the variance of the controls. In addition to dealing with heteroskedasticity, this solution can also deal with stochastic treatment effects heterogeneity (see Section \ref{Sec_Alternative_targets} for more details).

\subsubsection{Methods that exploit time-series variation} \label{Sec_N1_time_series}

Another alternative is to exploit the time-series variation. In such cases, we generally consider settings in which $T_0 \rightarrow \infty$, and we need to impose assumptions on the time-series of the errors (e.g., stationarity and weak dependence). Exploiting the time-series dimension allows for constructing inference methods that are typically valid with $N_0$ fixed. These methods typically only require models for the potential outcomes of the treated units, and allow for spatial correlation and richer heteroskedasticity in the cross-section. Therefore, these methods are complementary to the methods reviewed in Section \ref{Sec_N1_cross_section} in terms of the dimensions in which we need to impose strong assumptions and those in which we can allow for more flexibility. 

As in Section \ref{Sec_N1_cross_section}, we consider a setting with $N_1=T_1=1$ and no stochastic treatment effect heterogeneity, so that $\tau^\ast = \tau_{1,T}$ (we relax this assumption in Section \ref{Sec_Alternative_targets}).  If the errors $\{\epsilon_{1,t}\}_{t=1}^T$ are stationary, then testing $H_0:\tau^\ast=0$ is akin to testing the null hypothesis of no structural break at the end of the sample. More specifically, under the null of no effect and stationarity of $\{\epsilon_{1,t}\}_{t=1}^T$, we have that $Y_{1,t} - M_{1,t} = \epsilon_{1,t}$ for $t=1,\dots,T$ is stationary. \citet{hahn2017synthetic} suggest testing this implication using the end-of-sample instability test of \citet{andrews2003end} in the context of the synthetic control method, while \citet{ferman2019inference} propose to apply this test to make inferences in DiD applications with large $T_0$. The basic idea is to compare the post-treatment residual under the null $H_0:\tau^\ast=0$, $\hat\epsilon_{1,T}=Y_{1,T}-\hat{M}_{1,T}=Y_{1,T}(1)-\hat{M}_{1,T}=Y_{1,T}(0)-\hat{M}_{1,T}$, to the pre-treatment residuals $\{\hat\epsilon_{1,t}\}_{t\in \mathcal{T}_0}$ and reject if $\hat\epsilon_{1,T}$ is large relative to $\{\hat\epsilon_{1,t}\}_{t\in \mathcal{T}_0}$, where $\hat \epsilon_{1,t} = Y_{1,t} - \hat M_{1,t}$.\footnote{If we are interested in testing $H_0:\tau^\ast=c$, then the post-treatment residual (under the null) is $\hat\epsilon_{1,T}=Y_{1,T}-c-\hat{M}_{1,T}$.} For example, one can compute (two-sided) $p$-values for testing $H_0:\tau^\ast=0$ as

\begin{equation}
\hat{p}=\frac{1}{T}\sum_{t=1}^T\mathbf{1}\{|\hat\epsilon_{1,t}|\ge |\hat\epsilon_{1,T}|\}.\label{eq:p-value_time_series}
\end{equation}

The $p$-value in \eqref{eq:p-value_time_series} is asymptotically valid under two conditions. First, since $\hat\epsilon_{1,t}-\epsilon_{1,t}=\hat{M}_{1,t}-M_{1,t}$, the estimation error $\hat{M}_{1,t}-M_{1,t}$ needs to be negligible so that the difference between the estimated errors $\hat\epsilon_{1,t}$ and the true errors $\epsilon_{1,t}$ is negligible.\footnote{{See discussion in Appendix \ref{app_mis} for the case in which $M_{1,t}$ is misspecified. }}
Second, the true errors need to be stationary and weakly dependent so that the (infeasible) $p$-value based on the true errors, $p=T^{-1}\sum_{t=1}^T\mathbf{1}\{|\epsilon_{1,t}|\ge |\epsilon_{1,T}|\}$, is valid. Importantly, however, because this approach only exploits the time-series dimension of the problem, it does not require assumptions on the cross-sectional heteroskedasticity.

Many modern approaches for estimating $M_{1,t}$ involve high-dimensional estimation problems. A leading example is the synthetic control method where researchers are estimating a weight for each control unit based on the pre-treatment data, so that there are $N_0$ parameters and $T_0$ data points. In many applications, $T_0$ is small or moderate, while $N_0$ is comparable to or even larger than $T_0$. In such applications, the estimation error $\hat{M}_{1,t}-M_{1,t}$ can be substantial and render inference methods based on asymptotic approximations inaccurate. To address this challenge, \citet{chernozhukov2021exact}  propose a conformal inference method that has a double-justification: it is valid in finite samples when the untreated potential outcomes are exchangeable in the time-series dimension (e.g., when the potential outcomes are iid across time), and it is asymptotically valid when the data exhibit dynamics and time-series dependence as $T_0\rightarrow \infty$. 

The key idea of the conformal inference procedure is to estimate $\hat{M}_{1,t}$ using the data from all $T$ periods under the null hypothesis $H_0:\tau^\ast=0$. Estimation under the null hypothesis implies that $\{\hat\epsilon_{1,t}\}_{t=1}^T$ is exchangeable if the data are exchangeable, which implies that the method is exact in finite samples based on classical arguments for randomization tests \citep[e.g.,][]{hoeffding1952large,romano1990behavior}, that is $\mathbb{P}[\hat{p}\le \alpha]\le \alpha $ for any $(N,T)$. \citet{chernozhukov2021exact} show that estimation under the null is also crucial for a good finite sample performance with dependent (non-exchangeable) data. Intuitively, this is because estimation under the null ensures that all residuals are affected equally by the estimation error. By contrast, if $M_{1,t}$ is estimated using only pre-treatment data, then $\{\hat\epsilon_{1,t}\}_{t=1}^{T_0}$ may be too small compared to $\hat\epsilon_{1,T}$ due to overfitting, which can lead to substantial size distortions. \citet{cattaneo2021prediction} develop an alternative approach for dealing with the estimation error $\hat{M}_{1,T}-M_{1,T}$. Since their method is designed for constructing prediction intervals when the treatment effects are stochastic, we discuss it in Section \ref{Sec_Alternative_targets}.

Finally, when exploiting the time-series dimension, non-stationary data are ubiquitous and lead to non-standard behavior of common estimators of $M_{j,t}$. To this end, \citet{masini2021counterfactual_jasa,masini2022counterfactual_jbes} characterize the asymptotic properties of regression and Lasso estimators of $M_{1,t}$ when the data are non-stationary. For settings with $T_1=1$ and $T_0\rightarrow \infty$, their theoretical results justify a residual-based inference method as discussed above when the errors are stationary.

\begin{remark}

When errors are predictable (for example, when $\epsilon_{1,t}$ exhibits serial dependence), \cite{goncalves2024imputation}, show that it is possible to improve the mean squared error of existing estimators by adding a correction to $\hat \tau$ based on a prediction of $\epsilon_{1,T}$ using pre-treatment data (see also \cite{chernozhukov2021exact} and \cite{fan2022do} for related proposals). This adds a subtle decision on whether inference should be unconditional or conditional on past shocks. See \cite{alvarez2024_PUP} for a discussion on the validity of tests exploiting time-series variation when we consider them for conditional or unconditional inference.

\end{remark}

\subsubsection{Allowing for stochastic treatment effects heterogeneity} 
\label{Sec_Alternative_targets}

In Sections \ref{Sec_N1_cross_section} and \ref{Sec_N1_time_series}, we focused on methods that are valid under the assumption that there is no stochastic treatment effect heterogeneity. That is, the treatment effects $\tau_{j,t}$ are fixed parameters. This assumption is strong in many settings: it requires $Y_{j,t}(1)$ and $Y_{j,t}(0)$ only differ by a constant, so that the treatment only has a location-shift effect. In this section, we discuss alternatives for settings with stochastic treatment effect heterogeneity.

\paragraph{Alternative inferential goals:} As discussed in \cite{AF2025}, the methods discussed in Sections \ref{Sec_N1_cross_section} and \ref{Sec_N1_time_series} that are valid when there is no stochastic treatment effect heterogeneity might  also be valid for alternative inferential goals when there is stochastic treatment effect heterogeneity. As a first alternative, these methods would be valid for testing ``sharp'' null hypotheses of the type $H_0: \mathbb{P}[Y_{1,T}(1) = Y_{1,T}(0) + c]=1$. The main idea is that, under this null, we have the homogeneity in treatment effects these methods  require for inference on $\tau^\ast$. As a second alternative, one can think of inference on the \textit{realized} treatment effects. In this case, while we consider a setting in which treatment effects are stochastic, we consider inference conditional on the treatment effect of the treated units. \cite{AF2025} provide further discussion on the rationale for considering that as an inferential goal  and on the required assumptions on the errors and treatment effect heterogeneity for valid inference on the realized treatment effects. In particular, we would need that the assumptions on the errors in Sections \ref{Sec_N1_cross_section} and \ref{Sec_N1_time_series}  must be valid even once we condition on the treatment effects. This would be the case, for example, when treatment effects and untreated potential outcomes are independent, which might be an unreasonable assumption in some settings. Finally, recent work has considered the construction of prediction (as opposed to confidence) intervals in settings with few treated units and stochastic treatment effects \citep{cattaneo2021prediction,cattaneo2023uncertainty,chernozhukov2021exact}. The idea in this case is to construct interval-valued estimators that contain the \textit{stochastic} treatment effect with a probability greater than a given confidence level. This alternative inferential goal requires weaker assumptions on the dependence between treatment effects and untreated potential outcomes relative to the assumptions required for inference on the realized treatment effects. \cite{AF2025} show that  confidence intervals resulting from inversion of the tests discussed in Sections \ref{Sec_N1_cross_section} and \ref{Sec_N1_time_series} may be alternatively interpreted as prediction intervals that are valid under treatment effect heterogeneity and arbitrary dependence between treatment effects and untreated potential outcomes.

\paragraph{Sensitivity analysis.} If researchers are interested in $\tau^\ast$ and unwilling to change the target parameter, they can perform an analysis to assess the sensitivity of the results with respect to the distribution of the heterogeneous treatment effects, or the distribution of $Y_{1,T}(1)$. \cite{hagemann2020inference} introduces a valid inference method for a setting with a single treated cluster in an asymptotic framework where the number of observations within each cluster is large but the number of clusters is fixed. In \cite{hagemann2020inference}'s setting, if a normal approximation holds for the average of outcomes in each cluster, then a valid test of a null $\tau^\ast = c$ can be conducted, provided that the researcher provides an upper bound for $ \bar{\rho} \equiv \mathbb{V}[\bar{Y}_{1,T}(1)]/\min_{j \in \mathcal{J}_0}\mathbb{V}[\bar{Y}_{j,T}(0)]$, where $\bar{Y}_{j,T}(d)$ denotes the average potential outcome in cluster $j$, with the first cluster being the treated one.\footnote{\citeauthor{hagemann2020inference}'s results also cover the more general case where one of the control clusters has zero variance. In this case, a bound on the ratio between the variance of the average outcome in the treated cluster and the \emph{second} smallest variance among control clusters must be specified.} Note that \cite{hagemann2020inference}'s method can be alternatively recast as a sensitivity analysis: given a significance level $\alpha$, one can find the smallest-value of $\bar{\rho}$ compatible with not rejecting the null.

\subsection{Methods exploiting $N_1>1$} \label{Sec: N1>1}

The methods reviewed in Section \ref{Sec_N1} typically also work in settings with $N_1$ fixed, but greater than one. However, settings with $N_1 > 1$ provide other alternatives for inference. Since we have more variation in the data on potential outcomes under treatment, we do not necessarily have to rely on information from the control units or from the pre-treatment periods to learn about the errors on the treated. This allows us to accommodate heteroskedasticity and stochastic treatment effect heterogeneity. Therefore, we consider the case in which $\tau_{j,t}$ may be stochastic, and we focus on the target parameter $\tau^\ast$ defined in Equation \ref{eq_ATT}. 

While these methods can accommodate heteroskedasticity and stochastic treatment effect heterogeneity, they generally rely on assumptions that would not be required by the methods reviewed  in Section \ref{Sec_N1}. Moreover, these methods usually still rely on stronger assumptions than standard inference methods that are asymptotically valid in settings with many treated and many control units. Finally, an important practical consideration is that these methods may have low or even trivial power when $N_1$ is very small. Thus, from a power perspective, the methods presented in Section \ref{Sec_N1} might be preferable in some settings with $N_1>1$ when $N_1$ is very small.  See Appendix \ref{app_mc} for simulation evidence. An important challenge when comparing alternative inference methods in terms of power, though, is that alternative methods generally rely on non-nested sets of assumptions.

\subsubsection{Sign-changes \& Wild Bootstrap}
\label{Sec: sign_changes}

\paragraph{Idea and implementation of sign-changes \& relationship with wild bootstrap.} We start by considering randomization inference methods based on sign changes. To understand the main idea of these methods, consider a cross-sectional setting where $N_1$ is fixed and the estimator $\hat \tau$ can be written as $\hat \tau = \frac{1}{N_1} \sum_{j \in \mathcal{J}_1} \hat \tau_j$, with $\hat \tau_j \overset{p}{\rightarrow} \tau^\ast + (\tau_j - \tau^\ast) + \epsilon_j$ when $N_0 \rightarrow \infty$. This setting encompasses a comparison of means with  $N_1$ fixed and $N_0 \rightarrow \infty$. In this case, if we define  $\hat \tau_j = Y_j - \frac{1}{N_0} \sum_{j' \in \mathcal{J}_0} Y_{j'}$ for each  $j \in \mathcal{J}_1$, then $\hat \tau_j \overset{p}{\rightarrow} \tau_j + \epsilon_j$ when $N_0 \rightarrow \infty$, if we can apply a law of large number for the average errors of the controls. Recall also that a DiD setting with no variation in treatment timing can be recast as a comparison of means, so this would  be another example.
 
The main idea in this case is that, if we assume that  $(\tau_j - \tau^\ast) + \epsilon_j$ are independent across $j$ and symmetric about zero, then the asymptotic distribution of $(\hat \tau - \tau^\ast)$ would be invariant to the group of transformations $\mathcal{G} \equiv \{-1,1\}^{N_1}$, meaning that, for any $g = (g_1,g_2,\ldots, g_{N_1})' \in \mathcal{G}$, the asymptotic distribution of $\hat \tau^g(\tau^*) =\frac{1}{N_1} \sum_{j \in \mathcal{J}_1} g_j (\hat \tau_j - \tau^\ast)$ would not depend on the choice of $g$. Therefore,  we can apply the theory of randomization inference under approximate symmetry from \cite{Canay2017} (see also \cite{CaiCanay2023} for implementation details). To test the null $H_0: \tau^\ast = c$, we compute $\{\hat \tau^g(c)\}_{g \in \mathcal
{G}}$, and compare $(\hat \tau - c)$ to this randomization distribution. The p-value would be given by $\hat p = \frac{1}{|\mathcal{G}|} \sum_{g \in \mathcal{G}}{\mathbf{1}\{|\hat \tau -c | \leq |\hat \tau^g(c) | \}}$. Under the null, this p-value satisfies $\limsup_{N_0\to \infty} \mathbb{P}[\hat p \leq \alpha] \leq \alpha$, for any $\alpha \in (0,1)$. Therefore, asymptotically, this test controls  size.

Intuitively, under the null hypothesis that $\tau^\ast = c$, the asymptotic distribution of $(\hat \tau_1 - c, \ldots, \hat \tau_{N_1} - c)$ is centered around zero. If we also assume symmetry, its asymptotic distribution  is  invariant to sign changes. That is, $(\hat \tau_1 - c, \ldots, \hat \tau_{N_1} - c)$ has the same asymptotic distribution  as $(g_1(\hat \tau_1 - c), \ldots, g_{N_1}(\hat \tau_{N_1} - c))$ for any $g \in \mathcal{G}$. As a result, the probability that $\hat \tau$ is among the $k$ largest values of  $\{\hat \tau^g(c)\}_{g \in \mathcal{G}}$, would (asymptotically) be $k/\operatorname{dim}(\mathcal{G})$, which guarantees that the test controls  size. Under the alternative hypothesis $\tau^\ast > c$, however, each $\hat \tau_j - c$ tends to be positive. In this case, flipping signs at random typically reduces the magnitude of the average, so most $\hat \tau^g(c)$ will be smaller than the observed $\hat \tau - c$. This shift in distribution leads to small p-values, giving the test power against alternatives where $\tau^\ast \neq c$.

Note that assuming that $(\tau_j - \tau^\ast) + \epsilon_j$ is symmetric allows for heteroskedasticity in the error term $\epsilon_j$, and for \textit{stochastic}  treatment effects heterogeneity $(\tau_j - \mathbb{E}[\tau_j])$,  which was not generally allowed in the inference methods considered in Section \ref{Sec_N1}. However, these methods require assumptions that were not required by the inference methods considered in Section \ref{Sec_N1}, such as symmetry. This symmetry condition is often justified in settings in which outcomes $Y_j$ represent averages of individual level or time-series observations within unit $j$, under assumptions that allow us to rely on a CLT within each unit \citep{Canay2017}.\footnote{\cite{dias2021inferencemultivaluedheterogeneoustreatment} introduce a test for null hypotheses that a quantile $\kappa$ of $Y(1)$ equals that of $Y(0)$ for settings with few treated units. Their test is based on the idea that, under the null, $\mathbb{P}(Y_j \leq Q_{Y(0)}(\kappa)) = \kappa$ for $j \in \mathcal{J}_1$. Under the assumption that $Y(1)$ and $Y(0)$ are symmetric (which is slightly different from the symmetry assumption required in the sign-changes method), their test with  $\kappa=0.5$ would also be valid for the null $\tau^\ast = 0$. Their framework also allow for non-/semi-parametric models with multi-valued heterogeneous treatments and dynamic treatment effects. } Moreover, these methods do not allow for deterministic  treatment effects heterogeneity, as in this case $\tau_j - \tau^\ast$ would not be symmetric about zero for all $j\in \mathcal{J}_1$. {This is a more general feature of methods discussed in this section: they may allow for some stochastic heterogeneity, but, since there is only a finite number of treated units, there is not enough information to discern between nonstochastic location shifts and stochastic treatment effect heterogeneity. In contrast, methods in previous sections allowed for deterministic heterogeneity by \emph{extrapolating} from controls or pre-treatment periods.}

The idea of a test based on sign-changes is similar in spirit to a wild bootstrap \citep{CGM,MacKinnon2017,MacKinnon2018,Canay_wild_bootstrap}.\footnote{In a wild bootstrap, the distribution of $\hat{\tau}$ is approximated by generating bootstrap samples in which the estimated residuals are multiplied by random weights that have mean zero and unit variance (e.g., $\pm 1$ with equal probability).} Indeed, in Appendix \ref{app_equivalence} we show conditions in which sign-changes is asymptotically equivalent to a wild bootstrap \textit{with the null imposed} in settings with $N_1$ fixed and $N_0 \rightarrow \infty$.

\paragraph{Power issues.} A practical limitation of inference methods based on sign-changes is that they never reject the null if $\alpha<1/\operatorname{dim}(\mathcal{G})$, and thus have trivial power in this case. In the extreme case with $N_1=1$, there are only two possible transformations with $|\hat \tau^g| = |\hat \tau|$ for $g \in \{-1,1\}$, so the p-value is equal to one and the test has trivial power for any $\alpha$. Likewise, with $N_1=2$ we would only have two distinct values for $|\hat \tau^g|$, so $\hat p$ could only be equal to 0.5 or 1. For a test at the 10\% (5\%) level to have non-trivial power, i.e. for the rejection probability to be greater than zero, there must be at least five (six) treated units \citep{CaiCanay2023}.  Therefore, the alternatives reviewed in Section \ref{Sec_N1} might be preferred in settings with $N_1$ very small (even when $N_1>1$), due to power considerations.  See Appendix \ref{app_mc} for an illustration. Importantly, though, these methods rely on non-nested sets of assumptions. Therefore, we should be careful when contrasting alternative inference methods in this setting based on power.

\paragraph{Applications of sign-changes.} Inference methods based on sign-changes  have been considered in DiD settings with few treated and many control units when we have uniform treatment timing \citep{Canay2017}. In this case, we define $\epsilon_j$ as the post-pre average errors.  These methods have also been considered for matching estimators with few treated and many control units \citep{ferman_matching}. In this case, for $j \in \mathcal{J}_1$, we can define $\hat \tau_j$ as the difference between $Y_j$ and the average of its nearest neighbors. \cite{ferman_matching} provides conditions under which $\hat \tau_j \overset{p}{\rightarrow} \tau^\ast + (\tau_j - \tau^\ast) + \epsilon_j - \xi_j$ as $N_0 \rightarrow \infty$, where $\xi_j$ is the average of the errors of the nearest neighbors of treated unit $j$. If units are independent and the probability that two treated units share the same nearest neighbor goes to zero, then dependence between $\hat \tau_j$'s would go to zero when $N_0 \rightarrow \infty$. Therefore, the sign-changes test would be asymptotically valid when $N_1$ is fixed and  $N_0 \rightarrow \infty$ if $(\tau_j - \tau^\ast) + \epsilon_j - \xi_j$ is symmetric about zero. \cite{ferman_matching} considers finite-sample corrections to take into account that, in finite samples, treated units may share the same nearest neighbor. \cite{ferman_matching} also conjectures conditions under which the sign-changes test may be used in synthetic control applications with more than one treated unit. Randomization tests based on sign changes are also applicable, for example, in RCTs with few treated and many control units (so we consider an asymptotic approximation with $N_1$ fixed and $N_0 \rightarrow \infty$). Therefore, relative to the methods considered in Section \ref{Sec_N1_cross_section}, sign-changes randomization tests provide an alternative (in a model-based framework) that allows for stochastic treatment  effect heterogeneity at the expense of assuming symmetry.

\paragraph{Sign-changes with finite $N_1$ and $N_0$ validity.}
In some cases, it is also possible to consider a sign-changes method that is valid when both $N_1$ and $N_0$ are fixed. For example, \cite{Canay2017} discuss how their proposed sign-changes test may be adapted to settings with fixed $N_1$ and $N_0$, under additional assumptions. In this case, the idea is to cluster the units in $N_1$ disjoint groups, each one containing one treated and some of the control units. Suppose that the adopted estimator is ``linear'', in the sense that $\hat \tau = \frac{1}{N_1} \sum_{j \in \mathcal{J}_1} \tilde \tau_j$, with $\tilde{\tau}_j$ being an estimator of the treatment effect of unit $j$, using data from cluster $j$. Suppose each $\tilde{\tau}_j$ may be decomposed as  $\tilde \tau_j = \tau^\ast + (\tau_j - \tau^\ast) + \tilde \epsilon_j$, where $\tilde \epsilon_j$ is a function of the errors in the $j$-th cluster.  If the $(\tau_j - \tau^\ast) + \tilde \epsilon_j$ are independent and symmetric  about zero across $j$, the sign-changes test of the null $\tau^\ast = c$ based on $ \frac{1}{N_1}\sum_{j \in \mathcal{J}_1} g_j (\tilde \tau_j - c)$ and on the group of transformations $(g_1,\ldots,g_n) \in \mathcal{G}$, is valid with fixed $N_0$. 

One detail in this case is that the p-value would depend on the division of the units into disjoint groups. As a solution, \cite{CaiCanay2023} recommend users ``combine'' the conclusions from different partitions by relying on the strategies proposed by \cite{DiCiccio2020} in the context of a general hypothesis testing problem. Alternatively, we show in Section \ref{remark_avg} of Appendix \ref{app_equivalence} that another valid solution to this problem is to consider a number of different partitions, and then to construct an ``aggregate'' p-value across these partitions, following an idea similar to \cite{song2018orderingfree} and \cite{Leung}. By leveraging the special structure of the sign changes transformation and standard results on randomization tests, this alternative yields an inference procedure that is easier to implement and generally more powerful than the solutions in \cite{DiCiccio2020}. Moreover, as we remark in Appendix \ref{app_equivalence}, this aggregate p-value is asymptotically equivalent to the standard sign-changes test when $N_0 \rightarrow \infty$. Therefore, compared to the finite $N_0$ version of the sign-changes method proposed by \cite{Canay2017}, this alternative has the advantage of  not depending on the partition of the units, without requiring additional assumptions. At the same time, it is asymptotically equivalent to the standard implementation of the sign-changes method proposed by \cite{Canay2017} when $N_0$ is large.\footnote{ \cite{lau2025} provides an alternative approach that seeks to find a single partitioning scheme in order to maximize (local) power. This alternative is valid  with $N_0$ \emph{fixed} and (approximately) Gaussian $\tilde \epsilon_j$.}

\subsubsection{Behrens-Fisher solutions}

If we impose normality on the potential outcomes, then both a treatment–control cross-sectional comparison and a DiD design with uniform treatment timing reduce to testing for a difference in means between two groups with potentially unequal variances---a setting commonly known as the Behrens–Fisher problem \citep[e.g.,][]{Bakirov1998}.

In these cases, there are several alternatives available in the literature that could be used for settings with fixed $N_1$ and $N_0$. The  normality assumption may be justified as holding at least approximately in settings where each unit consists of an average of many observations, and it is plausible to consider a CLT for these within-unit averages. However, this justification for relying on normality approximations precludes the possibility of within-unit aggregate shocks.

\citet{Ibragimov2016} provide conditions on sample sizes and significance levels under which a degrees-of-freedom adjustment to an unequal variance $t$-test for the comparison of means of two populations is conservative when observations are independently normally distributed with possibly heterogeneous variances.\footnote{\cite{Bloom} apply \citeauthor{Ibragimov2016}'s (\citeyear{Ibragimov2016}) approach to an RCT with $N_1 = 11$ and $N_0 = 7$. They estimate effects for unit outcomes that consist of a before-after comparisons with a large number of time periods, so the Gaussian approximation for each individual observation is appropriate.} {More recently, \cite{potscher2023} extended the results of \cite{Ibragimov2016} to a general Gaussian linear model. They provide sufficient conditions for the existence of modified critical values that ensure the heteroskedasticity-robust t-test controls size in finite samples.\footnote{Their results also extend to some non-Gaussian settings, provided that the standardized errors of the model are spherically and symmetrically distributed with no point mass at zero.} Their critical values coincide with \citeauthor{Ibragimov2016} in the comparison-of-means problem for the ranges of sample sizes and significance values in \citeauthor{Ibragimov2016}, but are also computable for other sample sizes and significance values. \cite{potscher2024} shows that the sufficient condition in \cite{potscher2023} is in fact necessary for finite-sample size control of tests based on a t-statistic with heteroskedasticity-consistent standard errors}; and \cite{preinerstorfer2021} provides computational code to find the modified critical values.

In a  setting similar to the one from \cite{Ibragimov2016}, \cite{Hagemann2023_permutation} constructs a (generally conservative) permutation test for the null that the difference in means between the two populations is equal to some value $c$. His results cover different configurations of test statistics, sample sizes and significance levels than \citet{Ibragimov2016}.

Still considering alternative inference procedures, \cite{Ibragimov2010} consider settings in which the parameter of interest can be estimated separately in a finite number of independent clusters. In DiD designs, this may require considering coarser clusters, containing both treated and control units, similarly to the discussion in Section \ref{Sec: sign_changes}. In such settings, they provide conditions on sample size and significance levels for their procedure to be conservative. Under an additional assumption that estimators computed in each cluster have the same variance, we show in Appendix \ref{app_bester} that their procedure in DiD designs collapses to a modified version of the method of \cite{Bester2011}, which is exact regardless of sample size and significance level. Finally, \cite{hansen2024jackknife} introduces a jackknife variance estimator for linear regressions that is never downward-biased, while \cite{Hansen2025} studies its application to clustered DiD designs. He shows through simulations that confidence intervals that rely on this variance estimator can have better coverage when compared to confidence intervals based on cluster-robust standard errors.

Overall,  a common feature of the normality-based methods discussed in this section is that they become invalid or have trivial power when $\min \{N_1,N_0\} = 1$, meaning they are only viable in settings with $N_1,N_0>1$. Moreover,  even when $\min \{N_1,N_0\} > 1$, these methods can have lower power than those discussed in Section \ref{Sec_N1} if the number of treated units is very small. However, they can exhibit non-trivial power with small $N_1$ in settings where the alternatives from Section \ref{Sec: sign_changes} would have trivial power. See Appendix \ref{app_mc} for an illustration. We recall, though, that power comparisons for methods that rely on non-nested assumptions should be considered with caution.

\subsubsection{Dual justification: stronger assumptions with $N_1$ fixed \& weaker assumptions with $N_1,N_0 \rightarrow \infty$}

\label{Sec: dual}

A common theme for methods that are valid with fixed $N_1$ (whether with fixed $N_0$ or $N_0 \rightarrow \infty$) is that they rely on  stronger assumptions relatively to standard methods that are valid with large $N_1$ and $N_0$. This motivates inference methods that are valid under stronger assumptions when $N_1$ (or $N_1$ and $N_0$) is fixed and under weaker assumptions when $N_1,N_0 \rightarrow \infty$. One way to achieve that in methods that are based on randomization inference (such as those based on permutations or sign changes) is to consider studentized test statistics, as considered by, for example, \cite{Janssen1997}, Chapter 15 of \cite{Lehmann2005}, \cite{Chung2013}, \cite{DiCiccio2017}, \cite{MACKINNON2020435}, \cite{ferman_matching}, \cite{Canay_wild_bootstrap}, \cite{Bertanha2023} and \cite{DHaultfuille2024}. This is also a feature of the approaches considered  by \cite{Ibragimov2016} and \cite{Bester2011}, as it is well-known that $t$-tests with heteroskedasticity-robust (cluster-robust) standard errors are asymptotically valid under much weaker conditions when there are many treated and untreated units (clusters). Following a similar idea, \cite{Chaisemartin2022} propose a testing procedure in a DiD setting that is exact in finite samples under normality, homoskedasticity, and treatment effect homogeneity; and that remains asymptotically valid when these assumptions are relaxed.

\subsubsection{Higher-order improvements to standard asymptotics}

Another related set of approaches consists in constructing inference procedures that are valid under standard asymptotics as $N_0,N_1\rightarrow \infty$, while achieving better performance in finite samples. This includes corrections to standard error formulae with an aim to remove higher-order bias terms and downweigh high-leverage data points \citep{Mckinnon1985,Cribari12000,Cribari2004,Cribari2007,mackinnon2023cluster}, and approaches that, building on an idea originally due to \cite{Welch1951}, seek to compute improved critical values for $t$-tests by relying on a t-distribution with degrees of freedom chosen in a way such that moments of the ratio between the variance estimator and its population counterpart mimic those of a chi-squared distribution \citep{bell2002bias,Imbens2016,young2016improved,Hansen2025}. Studentized bootstrap methods that achieve higher-order improvements can also be placed in this category \citep{CGM,cameron2015practitioner,Djogbenou2019,mackinnon2023cluster}. It is important to note that the theoretical justifications of these approaches still rely on asymptotic approximations where $N_1 \to \infty$, which may be inaccurate if $N_1$ is small.

\subsection{Methods that exploit within-cluster Central Limit Theorems}
\label{Sec: exploiting CLT}

Here we discuss methods that exploit CLTs for averages of within-cluster observations. 
We consider a setting where $N_1=1$, but $T_0$ and $T_1$ are large. The target parameter is the ATT for the treated unit over the entire post-treatment period,
$\tau^\ast = \mathbb{E}\left[ T_1^{-1} \sum_{t \in \mathcal{T}_1}\tau_{1,t} \right]$, where $\tau^\ast$ may change with $T_1$, but we suppress such dependence for notational convenience. In this setting, the availability of many post-treatment periods allows for developing inference methods based on CLTs.\footnote{For example, methods exploiting large $T_1$ in panel data settings have been proposed by, for example,  \cite{li2017estimation,carvalho2018arco,li2020statistical,synthetic_did,masini2021counterfactual_jasa,masini2022counterfactual_jbes,li2023statistical,chernozhukov2024ttest}. Some of those methods would nest classical DiD as a special case.} Ideally, one would like to establish asymptotic normality results, such as $\sqrt{T}_1(\hat\tau -\tau^\ast)\overset{d}\rightarrow N(0,\sigma_\tau^2)$ as $T_1\rightarrow \infty$, where the asymptotic variance $\sigma_\tau^2$ can be consistently estimated.\footnote{Some estimators may not be asymptotically normal, even when $T_1\rightarrow \infty$ \citep{li2020statistical}.} However, there are at least three major challenges.

The first challenge is dealing with the error from estimating $M_{1,t}$. Since this term is rescaled by $\sqrt{T}_1$, it is not sufficient to just have a consistent estimator for $M_{1,t}$. Moreover, as discussed in Section \ref{Sec_N1_time_series}, some approaches for constructing $\hat{M}_{1,t}$ involve high-dimensional estimation problems. In such cases, the simple plug-in estimator $\hat \tau = \frac{1}{T_1} \sum_{t \in \mathcal{T}_1} (Y_{1,t} - \hat{M}_{1,t})$ may be biased, motivating bias correction procedures \citep[e.g.,][]{chernozhukov2024ttest}.

The second challenge is to estimate the asymptotic variance when there is serial correlation. The estimation error $\sqrt{T}_1(\hat\tau -\tau^\ast)$ typically contains terms like $\frac{1}{\sqrt{T_1}} \sum_{t \in \mathcal{J}_1} \epsilon_{1,t}$, which satisfy a CLT if $\{\epsilon_{1,t}\}$ is stationary and weakly dependent, $\frac{1}{\sqrt{T_1}} \sum_{t \in \mathcal{J}_1} \epsilon_{1,t}\overset{d}\rightarrow N(0,\sigma_\epsilon^2)$. A standard approach to estimate $\sigma_\epsilon^2$ when $\{\epsilon_{1,t}\}$ exhibits serial correlation is to use \citet{newey1987simple} variance estimators \citep[e.g.,][]{li2017estimation,carvalho2018arco}. However, these estimators can perform poorly when the number of periods is small or moderate. To overcome this challenge, \citet{chernozhukov2024ttest} propose an inference method that avoids the estimation of $\sigma_\epsilon^2$. The idea is to construct ``self-normalized'' test statistic---a test statistic in which the denominator and the numerator are proportional to $\sigma_\epsilon$, which thus cancels out. 

The third challenge is dealing with non-stationary data, which is ubiquitous in applications where exploiting the time series dimension is useful for inference. One approach for dealing with non-stationarity is to impose explicit time series models for the non-stationarity and analyze the properties of specific estimators under these models \citep[e.g.,][]{li2020statistical,masini2021counterfactual_jasa,masini2022counterfactual_jbes}. The drawback of this approach is that the resulting inference methods are not robust against violations of the underlying models of non-stationarity. Another complementary approach is to impose restrictions on the heterogeneity in the non-stationarity across units \citep{chernozhukov2024ttest}.

In addition to these challenges, there is a trade-off between allowing for stochastic and deterministic treatment effect heterogeneity. If the treatment effect sequence $\{\tau_{1,t}\}_{t\in \mathcal{T}_1}$ is deterministic, we can typically allow for arbitrary effect heterogeneity over time \citep[e.g.,][]{carvalho2018arco,chernozhukov2024ttest}.\footnote{As discussed in Section \ref{Sec_Alternative_targets}, the confidence intervals obtained using the methods that treat $\{\tau_{1,t}\}_{t\in \mathcal{T}_1}$ as deterministic can often be reinterpreted as prediction intervals if $\{\tau_{1,t}\}$ is stochastic. See \cite{AF2025} for further discussion.} If the treatment effects are stochastic, stationarity and weak dependence assumptions on $\{\tau_{1,t}\}_{t\in \mathcal{T}_1}$ are typically required \citep[e.g.,][]{li2017estimation,li2020statistical,chernozhukov2024ttest}, ruling out deterministic treatment effect heterogeneity. 

On the one hand, settings with many post-treatment periods allow for developing inference methods based on asymptotic normality, making these methods easy to implement and communicate. Moreover, some of these methods can relax assumptions required by methods discussed in Sections \ref{Sec_N1} and \ref{Sec_N1_time_series}, such as homoskedasticity and the absence of stochastic treatment effect heterogeneity. On the other hand, these methods rely on asymptotic approximations, which may be inaccurate in applications where $T_1$ is small or moderate.

\begin{remark}[Inference without a distance metric]
   In Appendix \ref{sec: subsampling}, we consider an alternative setting in which the number of observations within each treated cluster goes to infinity, but we do not have information on the dependence structure between observations (for example, when the number of individual-level observations within each treated cluster goes to infinity, but the distance metric along which dependence decays is unknown). In this case, under weak dependence assumptions on the errors, we may still have that the ATT estimator is asymptotically normal, and it is possible to conduct inference based on subsampling methods, while remaining agnostic to the dependence structure within clusters, as proposed by \cite{Leung}.  
\end{remark}

\section{Design-based inference} \label{Sec: design_based}

\subsection{Notation and sources of uncertainty}

An alternative approach is to consider design-based inference, in which (at least part of) the stochastic variation comes from the treatment assignment. This approach has a long tradition in the analysis of experiments (\citeauthor{Neyman1990}, \citeyear{Neyman1990} [1923]; \citeauthor{Fisher1992}, \citeyear{Fisher1992} [1926]). In design-based analyses, potential outcomes of the sample are often considered as fixed, $\{Y_j(1),Y_j(0) \}_{j=1}^N$, and uncertainty would come from the treatment assignment $\{D_j\}_{j=1}^N$.\footnote{To be consistent with the notation from Section \ref{Sec: design_based}, we continue to consider $\{Y_j,D_j\}_{j=1}^N$ as the sample, which differs from \cite{Abadie_finitepop}, who let $N$ be the number of observations in the finite population.} The target parameter can be the sample average treatment effect (SATE), $\tau_{\mbox{\tiny SATE}} \equiv \frac{1}{N} \sum_{j=1}^N (Y_j(1) - Y_j(0))$. If treatment is randomly assigned, then the standard difference in means estimator, $\hat \tau = \frac{1}{\sum_{j=1}^N D_j}\sum_{j=1}^N Y_j D_j - \frac{1}{N - \sum_{j=1}^N D_j}\sum_{j=1}^N Y_j (1-D_j)$ would be unbiased for the SATE, in this framework in which the only source of uncertainty comes from the assignment of treatment \citep[Chapter 6]{Imbens_Rubin_2015}. Alternatively, we may consider the sample $\{Y_j,D_j\}_{j=1}^N$ as being drawn from a larger (finite or infinite) population, and define the target parameters as the Population Average Treatment Effect (PATE). In this case, uncertainty quantification should account for both randomness in the assignment mechanism, as well as sampling.

Note that the focus in such settings is generally on parameters related to  average treatment effects given the realized potential outcomes of the sample (or of a finite population), while in Section \ref{Sec: model based} the focus was on parameters related to the average treatment effects on the treated over different realizations of the potential outcomes.\footnote{An exception is \cite{Rambachan_designbased}, who also define the expected average treatment effect on the treated, as an analog of the ATT in model-based settings. Still, this target parameter is also defined conditional on the realization of the potential outcomes, which differ from the target parameter in model-based approaches.}  Therefore, from a conceptual perspective, a decision on whether to focus on model-based or design-based approaches for inference should reflect the target parameter of interest. 
From a pragmatic perspective, design-based approaches might be preferable when there is knowledge about the treatment assignment mechanism \citep{roth2023what}.  An extreme example is when the treatment is randomly assigned, in which case knowledge of the assignment mechanism can be exploited to construct exact inference procedures. Design-based approaches are also helpful in settings in which modeling the outcome process is difficult. This is especially useful in settings where it is difficult to posit the dependence structure underlying sampling uncertainty, as design-based procedures remain agnostic about these \citep{Barrios2012,Adao2019}. More recently, design-based approaches have also been considered in natural experiments, including in settings where treatment assignment probabilities are not equal across units \citep{Rambachan_designbased}.

\subsection{Asymptotic Inference in Design-Based Settings}

Inference based on $t$-statistics and asymptotic normal approximations has been proposed for testing hypotheses about the SATE or PATE in a variety of design-based settings, including cross-sectional treatment–control comparisons (linearly adjusted for controls or not) \citep{Abadie_finitepop,Abadie2022}, as well as other quasi-experimental designs \citep{Adao2019,Athey2022,Roth2023,Rambachan_designbased}. However, these methods rely on asymptotic approximations that require both the number of treated and control units to diverge. As a result, such methods may perform poorly when the number of treated units is small, for reasons similar to those discussed in Sections \ref{Sec: example} and \ref{Sec: model, parameters of interest, and estimators}.

\subsection{Randomization tests}

\subsubsection{Randomization tests in experiments and beyond}

In settings where the treatment assignment mechanism is known (e.g., RCTs), randomization tests constitute a natural alternative for conducting design-based inference \citep{fisher1949design,Imbens_Rubin_2015,young_QJE}. Randomization tests are exact in finite samples for testing sharp null hypotheses, such as $H_0: Y_j(1)=Y_j(0)$ for all $j$, and are therefore well-suited for applications with a small number of observations (or with few treated observations). They are valid even when there is only a single treated unit. To illustrate the main idea of randomization tests, consider the simple difference-in-means estimator.
\begin{eqnarray}
\hat \tau = \frac{1}{\sum_{j=1}^N D_j}\sum_{j=1}^N Y_j D_j - \frac{1}{N - \sum_{j=1}^N D_j}\sum_{j=1}^N Y_j (1-D_j) = \\ \frac{1}{\sum_{j=1}^N D_j}\sum_{j=1}^N Y_j(1) D_j - \frac{1}{N - \sum_{j=1}^N D_j}\sum_{j=1}^N Y_j(0) (1-D_j).   \end{eqnarray}

In an RCT, the distribution of the treatment assignment vector $(D_1,\dots,D_N)$ is known, so that if we knew both potential outcomes for all $j=1,...,N$, we would know the distribution of $\hat \tau$. Sharp null hypotheses such as $H_0: Y_j(1)=Y_j(0)$ for all $j$ allow us to compute both potential outcomes for each unit as $Y_j=Y_j(1)=Y_j(0)$, so that the distribution of $\hat \tau$ is known under the null. For example, in a completely randomized experiment with $N_1$ treated and $N_0$ control units, this distribution can be computed by recalculating $\hat \tau$ under each of the ${N\choose N_1}$ possible treatment assignments with $N_1$ treated units. Hypotheses tests can then be conducted by comparing the actual estimate to the quantiles of this randomization distribution.

Randomization tests have also been considered in a variety of non-experimental settings. Some examples include the synthetic control literature \citep{abadie2010synthetic,FirpoPossebom+2018,lei2024inference}, panel data settings with staggered treatment adoption \citep{Shaikh2021}, observational studies under a selection on observables assumption \citep{Rosenbaum2002}, instrumental variable designs \citep{Rosenbaum1996,Rosenbaum2002,Imbens2005}, and regression discontinuity designs \citep{CattaneoFrandsenTitiunik2015,Cattaneo2017,Bugni2021}. Such tests have also been considered in settings with non-random exposure to exogenous shocks \citep{Borusyak2023}, for which shift-share designs are a particular case \citep{Alvarez_shift_share}.

\subsubsection{Issues with randomization tests with few treated units}
\label{sec:issues_randomization_tests}
In the following, we highlight two important points regarding the use of randomization tests that are particularly relevant when we consider settings with few treated units. 

\paragraph{Testing a sharp null.} An important point to notice is that this finite-sample justification of randomization tests does not consider testing  null hypotheses regarding the SATE or PATE. Therefore, we may have that $\tau_{\mbox{\tiny SATE}}=0$, but a permutation test would reject at a rate greater than $\alpha$. This could happen, for example, if $\frac{1}{N} \sum_{j=1}^N Y_j(1) =  \frac{1}{N} \sum_{j=1}^N Y_j(0)$, but $\frac{1}{N} \sum_{j=1}^N (Y_j(1) - \bar Y(1))^2 >  \frac{1}{N} \sum_{j=1}^N (Y_j(0) - \bar Y(0))^2$, where $\bar Y(d) = \frac{1}{N}\sum_{j=1}^N Y_j(d)$ (that is, treatment does not affect the average of the potential outcomes, but affects their variances).

Another way to justify the use of permutation tests is to consider a test for the null $H_0: \tau_{\mbox{\tiny SATE}}=0$, but assume that $Y_j(1) = Y_j(0) + c$ for all $j$, for a constant $c$. Under this assumption, we have that the null $H_0: \tau_{\mbox{\tiny SATE}}=0$ implies $Y_j(1) = Y_j(0)$ for all $j$, which is the main building block to show that the permutation test is valid when we know the distribution of treatment assignment. Assuming that treatment effects are homogeneous in this finite-population setting is similar in spirit to assuming that there is no stochastic treatment effect heterogeneity in a model-based setting (e.g., \cite{conley20211inference}). This should not be surprising, given the asymptotic equivalence between a permutation test and the method proposed by \cite{conley20211inference} in model-based settings. This highlights that, whether we are in  design- or model-based settings, permutation tests rely on the same kind of restrictions on the treatment effect heterogeneity for exact validity. 

Similarly to the discussion in Section \ref{Sec: dual}, an interesting feature of randomization tests of sharp nulls in design-based settings is that, by properly studentizing the test statistic, it is possible to construct tests that are exact in finite samples for the sharp null, and also asymptotically valid (albeit generally conservative) for a weaker null hypothesis. This has been considered, for example, in randomized experiments \citep{Wu2020,Bugni2018,Young2024}, and shift-share designs \citep{Alvarez_shift_share}.


\paragraph{Unconditional vs conditional inference.} Randomization tests are valid for \textit{unconditional} inference. Now consider a setting in which we observe a characteristic $W_j \in \{0,1\}$ for the units in our sample (so that $\{Y_j(1),Y_j(0),W_j\}_{j=1}^N$ is treated as fixed). It might be that after running the experiment, the experimenter observes that there is an important imbalance in terms of the average of $W_j$   between the treated and the control group. Let $\bar W_1$ ($\bar W_0$) be the average of $W_j$ for the treated (control) group. In this case, while the randomization test would remain valid for unconditional tests of the sharp null, it may cease to be valid conditional on the fact that we had an imbalance in $(\bar W_1,\bar W_0)$. In this case, an alternative for valid conditional inference would be to consider only permutations with the same imbalance $(\bar W_1,\bar W_0)$ \citep{Hennessy2016}. Whether to  use unconditional or conditional inference in such settings has been subject to a longstanding debate \citep[see][and references therein]{Mutz2019,Johansson2022}.  We highlight that this discussion is particularly relevant in settings with few treated units, which is the focus of our survey. In such settings, the probability of having relevant imbalances is higher than with many treated and many control units. In particular, if we consider a setting with $N_1=1$ and $W_j \in \{0,1\}$, then we would have that $\bar W_1 \in \{0,1\}$, implying imbalance in all realizations of the treatment assignment whenever $W_i$ is not constant across units (i.e., whenever both values 0 and 1 are present in the sample). {In our view, conditional inference may be more appropriate, particularly when there are reasons to believe that conditioning on realized imbalances in the data is likely to induce size distortions. For example, this may occur in state-level settings when the treated state is much smaller than the control states (see Footnote \ref{Footnote: unconditional CT} for a related discussion). }

\section{Recommendations for practice}
\label{recommendations}

Whether we consider model-based or design-based settings, making inference about target parameters related to average treatment effects in applications with few treated units is inherently difficult, due to the limited information available on treated potential outcomes. Standard inference methods relying on asymptotic approximations justified by many treated and many control units can be unreliable and misleading when the number of treated units is small. At the same time, methods that are specifically designed for and valid in settings with few treated units typically rely on stronger assumptions than approaches that are valid with many treated units, or they focus on alternative inferential targets to compensate for the underlying lack of information.

The optimal choice of method is highly context-specific, and applied researchers will often have to choose between alternatives that rely on non-nested sets of assumptions. The structure of this survey is intended to help applied researchers navigate these trade-offs. When choosing an appropriate inference method, researchers should first decide between model-based and design-based approaches. These two approaches are conceptually distinct, and the choice between them should be guided by the target parameter of interest and by the notion of uncertainty that is relevant in the application at hand.

Given this decision, data restrictions may further limit the set of available options. For example, in settings with $N_1=1$, the methods discussed in Section \ref{Sec: N1>1} cannot be used. Even when $N_1>1$, some of the methods discussed in Section \ref{Sec: N1>1} may have trivial power when $N_1$ is very small, and therefore may not be informative in practice. Likewise, the methods discussed in Section \ref{Sec_N1_time_series}, which extrapolate information from pre-treatment periods, may be inappropriate when the pre-treatment period is short.

Still, in many settings applied researchers will have a menu of alternative inference methods to choose from even after ruling out those that are infeasible or uninformative due to data limitations. A natural example is the case with $N_1>1$. In such settings, the methods described in both Sections \ref{Sec_N1} (methods for settings with $N_1=1$) and \ref{Sec: N1>1} (methods for settings with $N_1>1$) are typically valid, and the choice among alternatives may involve non-trivial trade-offs related to the required assumptions (which are often non-nested across methods) and to power considerations.

In such cases, we recommend that researchers prioritize methods based on the empirical credibility of the required assumptions in the application at hand. Importantly, we encourage researchers to embrace the difficulty of the underlying inference problem and to be transparent about the additional assumptions required to achieve inferential validity in applications with few treated units, conditional on the chosen method. If multiple alternatives rely on assumptions that appear reasonable in the application at hand, the choice among them can be guided by power considerations, as discussed in Appendix \ref{app_mc}. Another alternative is to present p-values for all methods that rely on assumptions that are arguably reasonable for the application at hand. In this case, considering the maximum p-value across a set of inference methods guarantees a conservative test (i.e., size control) if at least one of the included methods yields a valid p-value under the null. An important caveat, however, is that in settings with few treated units some methods may be overly conservative under the null, and including such methods in the pool of options would produce an underpowered test.

\section{Directions for future research}
\label{directions}

The study of inferential approaches in causal settings with few treated units is a broad topic, and we believe there are several interesting venues for future research. On the one hand, we have seen that different methods typically rely on non-nested sets of assumptions; on the other hand, different approaches entail different power functions. As a consequence, we believe that thinking more formally about the trade-offs between validity and power, e.g. by explicitly accounting for the possibility of mis-specification in the theory, can lead to interesting insights. Relatedly, since the methods designed for settings with few treated units inherently rely on stronger assumptions that standard methods that are valid with many treated and many control units, the development of systematic sensitivity tools to existing inferential approaches can aid researchers in assessing the robustness of the conclusions of these tests to violations of the underlying distributional assumptions. We also remark that, in settings with a large number of control observations, it is usually possible to  consistently estimate the distribution of $\epsilon_{i,t}$ for the control units, and that might be used to  assess the validity of \emph{some} of the assumptions required by these methods.  We believe that an interesting venue of future research would be to think systematically about such tests. 

Finally, we observe that nonparametric causal inference analyses with continuous treatments can also incur a ``few-treated-units'' problem, as in several situations the number of observations within a given bandwidth of a support point of the treatment dosage can be quite small. Therefore, we see the extrapolation of the methods analyzed in this survey to such settings as another interesting avenue for further research.

\setlength{\bibsep}{0pt}
\bibliographystyle{apalike}
\bibliography{references}
\appendix

\section{Inference with disaggregated within-cluster data}
\label{app_disag}

It is common to have situations in which researchers have access to disaggregated data. For example, they might have access to data on individuals $i$ for different states $j$ and time periods $t$, while the treatment is at the $j \times t$ level. In such cases, aggregating the data at the $j \times t$ level is useful to address concerns regarding the within $j \times t$ correlation (though not other types of correlations). In the main text, the inference methods did not explicitly exploit variation from within $j \times t$ cells, so they would be essentially the same, whether we have aggregated or disaggregated data (indeed, when we have disaggregated data, most of the methods we have reviewed can be implemented by first aggregating the data at $j \times t$ level, and then applying the method). In this Appendix, we discuss some specific issues and possibilities that arise when we have individual-level data.

\subsection{Inference conditional on cluster-level shocks} \label{sec_shocks} In settings with cluster-level treatment assignment and many individual-level observations for each cluster, a possibility in model-based analyses is to  view uncertainty as only coming from the sampling of individuals within cluster, while conditioning on cluster-level aggregate shocks. To understand this idea, consider a simplified setting with two clusters: $j=1$ (treated) and $j=2$ (control) and many individual-level observations indexed by $i$ within each group. Let $Y_{ij}(0) = \omega_j + \epsilon_{ij}$ and $Y_{ij}(1) = \tau^\ast + Y_{ij}(0)$, where $\epsilon_{ij}$ is a mean-zero error iid across $i$ and $j$ and  independent from the aggregate shocks  $\omega_j$.  If we conduct inference using disaggregated data and heteroskedasticity-robust standard errors (ignoring aggregate shocks), this yields valid inferences on the target parameter $\tau^\ast +  \omega_1 - \omega_2$, conditional on the aggregate shocks $\omega_j$.  However, the issue with this approach is that $\tau^\ast +  \omega_1 - \omega_2$ does not have a clear causal interpretation within our potential outcomes framework. \cite{roth2023what} consider this idea  in DiD settings. As they recognize, conditioning on aggregate-level shocks would generally imply violation in the parallel trends assumption, so they recommend coupling this approach with bounding exercises, as proposed by \cite{manski2018right} and \cite{rambachan2023credible}.

\subsection{Conducting inference with few clusters while accounting for within-cluster dependence without a distance metric}

\label{sec: subsampling}
The methods discussed in Section \ref{Sec: exploiting CLT} relied on estimating the dependence structure between observations. This was possible because a metric along which the dependence between observations was assumed to decay -- in that case, time -- was naturally available. In other settings, however, treated observations may be arranged in clusters where no natural metric is available. For example, when we have many individual-level observations within each cluster.  For these cases, \cite{Leung} proposes a general inference method that remains agnostic about the dependence structure between observations. His approach remains valid even in settings with a single treated and a single control cluster, provided that the number of observations in each cluster ($N_j$) is large, and the conditions for the validity of a CLT on the within-cluster averages hold.  Relatively to the methods from Section \ref{Sec_N1_cross_section} and \ref{Sec_N1_time_series}, this approach requires access to individual-level data, and it requires the use of a CLT within cluster (so it precludes, for example, cluster-level shocks, which are allowed by these other methods).\footnote{The discussion in Appendix \ref{sec_shocks} on inference conditional on aggregate shocks applies to this setting.
}

In a comparison of means setting with one treated and one control cluster, \citeauthor{Leung}'s approach to inference relies on a standard two-sample $t$-test statistic, with the average in each cluster being computed by randomly resampling $R_j$ units with replacement from each cluster $j$. Provided that $R_j \to \infty$ with $R_j/N_j \to 0$, the results in \cite{Leung} ensure the test statistic converges in distribution to a standard normal, which enables researchers to construct tests with asymptotic validity. Importantly, the procedure remains agnostic about the dependence structure within each cluster (except for assuming that the dependence is such that a CLT within clusters is valid, which requires some sort of weak dependence within clusters). Intuitively, this is due to the restriction that  $R_j/N_j \to 0$, which ensures that resampled draws are approximately independently distributed, thus ensuring convergence to a standard normal even in the presence of dependence. Indeed, as argued by \citeauthor{Leung}, the restriction that $R_j/N_j \to 0$ may be seen as a price to pay in order to be agnostic about the dependence structure of observations, since we effectively ``lose'' observations by working with $R_j << N_j$ draws. Presently, there is no general method to choose $R_j$, though \cite{Leung} provides some guidance in specific settings.\footnote{The choice of $R_j$ is subject to a trade-off between statistical power and size control: a larger $R_j$ increases the power of the test, though possibly at the cost of a poorer quality of the normal approximation.}

\section{Interpreting $M_{j,t}$ under misspecification.}
\label{app_mis}

{In the examples in Section \ref{examples_M}, the mean-predictor $M_{j,t}$ is typically motivated as a conditional expectation of the missing potential outcome $Y_{j,t}(0)$ given the available information. In situations where one suspects that this conditional expectation may be misspecified,} $M_{j,t}$ can be alternatively interpreted as pseudo-true mean predictor and $\epsilon_{j,t} \equiv Y_{j,t}(0) - M_{j,t}$ as the pseudo-true error. For a given estimator $\hat M_{j,t}$, we can define $M_{j,t}$ as the probability limit of $\hat M_{j,t} $  \citep[e.g.,][]{cattaneo2021prediction,chernozhukov2021exact,goncalves2024imputation,alvarez2024_PUP}, so that $\hat M_{1,t} -  M_{1,t}$ is asymptotically negligible by construction. In this case, the difference from the correctly-specified setting lies in that the assumptions required for identification and inference, such as stationarity and weak dependence, would be required for the pseudo-true error $\epsilon_{1,t} \equiv Y_{1,t}(0) - M_{1,t}$. For example, consider a DiD setting with large $T_0$ and fixed $N_0$, where $Y_{j,t}(0) = \mu_j + \lambda_t + \nu_{j,t}$, and  define $$\hat M_{1,t} = \frac{1}{T-1} \sum_{t' \neq t} Y_{1,t'} + \frac{1}{N_0} \sum_{j \in \mathcal{J}_0} \left[Y_{j,t} - \frac{1}{T-1} \sum_{t' \neq t } Y_{j,t'} \right].$$
Assuming that $\frac{1}{T-1} \sum_{t' \neq t} \nu_{j,t'} \overset{p}\rightarrow 0$ when $T \rightarrow \infty$, we have $\hat M_{1,t} \overset{p}\rightarrow  M_{1,t} = \mu_j + \lambda_t + \bar \nu_t$, where $\bar \nu_t  = \frac{1}{N_0} \sum_{j \in \mathcal{J}_0} \nu_{j,t}$, implying that $\epsilon_{1,t} = \nu_{1,t}- \bar \nu_{t}$.\footnote{Note that, for some $t$, $\hat M_{1,t}$ will depend on outcomes $Y_{1,t'}$ from the treated periods. However, since we are considering a setting in which $T_1$ is fixed, while $T_0 \rightarrow \infty$, the treatment effects $\tau_{1,t'}$, for $t' \in \mathcal{T}_1$, will not affect $M_{1,t}$. } In this case, the methods discussed in this section would be valid if $\epsilon_{1,t} = \nu_{1,t}- \bar \nu_{t}$  is stationary and weakly dependent. In other settings,  the pseudo-true error might involve more complex terms, making the stationarity and weak dependence assumptions harder to justify. Therefore, these conditions need to be analyzed on a case-by-case basis (see \cite{alvarez2024_PUP} for other examples).

\section{Appendix to Section \ref{Sec_N1}}
\label{app_secn1}
\subsection{An exact version of \cite{conley20211inference}}
\label{app_exact}

\subsubsection{Without covariates}

In the model-based setting of Section \ref{Sec: model based}, \cite{conley20211inference} consider the following model for untreated potential outcomes,
\begin{equation}
\label{eq_model_twfe} Y_{jt}(0) = \lambda_t + \mu_j + \epsilon_{jt} \, .
\end{equation}

In an asymptotic framework where $N_1$ is fixed and $N_0$ is large, \cite{conley20211inference} show that the two-way fixed effects estimator of $\beta$ based on
\begin{equation}
\label{eq_twfe_est}
Y_{jt} = \lambda_t + \mu_j + \beta D_{jt}  + e_{jt}\, ,
\end{equation}
is such that 
\begin{equation}
    \label{eq_ct_limit}
\hat{\beta} -\frac{1}{N_1 T_1}\sum_{j \in \mathcal{J}_1} \sum_{t \in \mathcal{T}_1} \tau_{j,t} \overset{p}{\to}  \frac{1}{N_1}\ \sum_{j \in \mathcal{J}_1}  (\bar{\epsilon}_{j,1} - \bar{\epsilon}_{j,0}) \, ,
\end{equation}
where $\bar{\epsilon}_{j,d} = \frac{1}{T_d}\sum_{t \in \mathcal{T}_d} \epsilon_{j,t}$. \cite{conley20211inference} propose a resampling procedure that approximates the limiting distribution in \eqref{eq_ct_limit} by resampling $N_1$ differences of average residuals $(\overline{\hat{e}}_{j,0} - \overline{\hat{e}}_{j,1})$ \emph{with replacement from the control group}, and then uses these resampled residuals to compute an approximation to \eqref{eq_ct_limit}. Under the ``distributional'' parallel trends assumption:

\begin{equation}
    \label{app_pt_strong}
\overline{{\epsilon}}_{j,1}- \overline{{\epsilon}}_{j,0}  \text{ is iid over } j \in \mathcal{J}_0\cup \mathcal{J}_1 ,
\end{equation}
their procedure correctly recovers the nonstandard limiting distribution.

In their paper, \cite{conley20211inference} discuss an alternative inference procedure that resamples differences of average residuals computed under the null hypothesis and draws these \emph{without replacement from both the treatment and the control group} to approximate the limiting distribution \eqref{eq_ct_limit}. As they  argue, this procedure is asymptotically equivalent to their main resampling algorithm, since, with $N_0$ increasing and $N_1$ small, both the probability that a unit from the treatment group be selected in the resampling and the difference between null-imposed and standard residuals in the control group are arbitrarily small. In this appendix, we show that a small modification to this alternative resampling procedure -- namely, to use a ``long'' formula (including the errors of the controls) to approximate \eqref{eq_ct_limit} -- is sufficient to produce an exact test of the sharp null 
\begin{equation}
    \label{eq_sharp_ct}
\mathbb{P}\left[\frac{1}{T_1}\sum_{t \in \mathcal{T}_1}Y_{j,t}(1) = \frac{1}{T_1}\sum_{t \in \mathcal{T}_1}{Y}_{j,t}(0) + c\right]=1 \, ,\quad  j \in \mathcal{J}_1\,,
\end{equation}
under the strong parallel trends assumption,
\begin{equation}
\label{ass_pt_2}
\frac{1}{T_1}\sum_{t \in \mathcal{T}_1} Y_{j,t}(0) - \frac{1}{T_0}\sum_{t \in \mathcal{T}_0}Y_{j,t}(0) \text{ is iid over } j \in \mathcal{J}_0\cup \mathcal{J}_1 \, ,
\end{equation}
which, given the model \eqref{eq_model_twfe}, is equivalent to \eqref{app_pt_strong}.

Our proposed resampling procedure goes as follows. Suppose we are willing to test the null $\frac{1}{N_1 T_1} \sum_{j \in \mathcal{J}_1} \sum_{t \in \mathcal{T}_1} \tau_{j,t} = c$.  The researcher first estimates \eqref{eq_twfe_est} without imposing any restrictions, and stores the estimator $\hat{\beta}$. She then estimates \eqref{eq_twfe_est} by imposing the null $\beta = c$ and stores the null-imposed residuals  $\{\tilde{e}_{jt}\}_j$. The researcher then approximates the distribution of $\hat{\beta} - c$ under the null by considering the set of permutations $\Pi$ on $\{1,\ldots, N\}$, and computing, for each $\pi \in \Pi$, the approximation:
\begin{equation}
\label{eq_avg_pi}
\tilde{\beta}_\pi = \frac{1}{N_1}\sum_{j \in \mathcal{J}_1} (\overline{\tilde{e}}_{\pi(i),1} - \overline{\tilde{e}}_{\pi(i),0}) - \frac{1}{N_0}\sum_{j \in \mathcal{J}_0} (\overline{\tilde{e}}_{\pi(i),1} - \overline{\tilde{e}}_{\pi(i),0}) \, .
\end{equation}
    
    She then rejects the null if $(\hat{\beta} - c)$ is in the tails of $\{\tilde{\beta}_\pi \}_{\pi \in \Pi}$.

    We note that, with $N_1$ fixed and $N_0\rightarrow \infty$, the permutation distribution $\{\tilde{\beta}_\pi \}_{\pi \in \Pi}$ (asymptotically) coincides with \citeauthor{conley20211inference}'s two approaches to resampling, as both the average across $\mathcal{J}_0$ in \eqref{eq_avg_pi} and the contribution of the treated units in computing the residuals for the approximation become negligible. Moreover, we note that, for fixed $N_0$ and $N_1$, \eqref{eq_avg_pi} may be rewritten as:
    \begin{eqnarray*}
\tilde{\beta}_\pi &=&  \frac{1}{N_1}\sum_{j \in \mathcal{J}_1} \left(\frac{1}{T_1}\sum_{t \in \mathcal{T}_1} Y_{\pi(i),t} - \frac{1}{T_0}\sum_{t \in \mathcal{T}_0}Y_{\pi(i),t} - c D_{\pi(i)}\right) \\
&&- \frac{1}{N_0}\sum_{j \in \mathcal{J}_0} \left(\frac{1}{T_1}\sum_{t \in \mathcal{T}_1} Y_{\pi(i),t} - \frac{1}{T_0}\sum_{t \in \mathcal{T}_0}Y_{\pi(i),t} - c D_{\pi(i)}\right) \, ,
\end{eqnarray*}
which coincides with the reference permutation distribution of a standard permutation test. It then follows from the classical theory of randomization tests \citep[Chapter 15 of ][]{Lehmann2005,Canay2017} that, under assumption \eqref{ass_pt_2}, one can use  $\{\tilde{\beta}_\pi\}_{\pi \in \Pi}$ to construct an exact test of the sharp null \eqref{eq_sharp_ct}. This is because the distribution of the test statistic is invariant to permutations under the null \eqref{eq_sharp_ct} if \eqref{ass_pt_2} holds.

\subsubsection{With covariates}
\label{remark_cov_ct}
    \cite{conley20211inference} also consider a version of their test to the case with covariates. In this case, they consider the following model for untreated potential outcomes as follows:

    $$\frac{1}{T_1}\sum_{t \in \mathcal{T}_1}Y_{j,t}(0) - \frac{1}{T_0}\sum_{t \in \mathcal{T}_0}Y_{j,t}(0)   =\phi'X_j + \epsilon_j$$

    Under the assumption that the $\{\epsilon_i\}$ are iid and independent of the $\{X_i\}$, the results in \cite{conley20211inference} allow us to construct a test for the null $\tau^* = c$ that is asymptotically valid when $N_0 \to \infty$ (absent stochastic treatment effect heterogeneity), though this approach would not be exact in finite samples. 
    
    We consider a modification of their approach that is exact in finite samples for testing the sharp null that $\mathbb{P}\left[\frac{1}{T_1}\sum_{t \in \mathcal{T}_1}Y_{j,t}(1) = \frac{1}{T_1}\sum_{t \in \mathcal{T}_1}{Y}_{j,t}(0) + c\right]=1$ for every $j \in \mathcal{J}_1$, under the assumption that the pairs $\left(\frac{1}{T_1}\sum_{t \in \mathcal{T}_1}Y_{j,t}(0) - \frac{1}{T_0}\sum_{t \in \mathcal{T}_0}Y_{j,t}(0), X_j\right)$ are iid across $j \in \mathcal{J}_0 \cup \mathcal{J}_1$. Consider the coefficient $\hat \beta$ on $D_i$ of a regression of $\frac{1}{T_1}\sum_{t \in \mathcal{T}_1}Y_{i,t} - \frac{1}{T_0}\sum_{t \in \mathcal{T}_0}Y_{i,t}$ on $D_i$ and $X_i$. Notice that, under our stated sharp null:

    $$\hat \beta  -  c  = \frac{\sum_{i=1}^N (D_i - \tilde \delta' X_i) \left(\frac{1}{T_1}\sum_{t \in \mathcal{T}_1}Y_{i,t}(0) - \frac{1}{T_0}\sum_{t \in \mathcal{T}_0}Y_{i,t}(0)\right)}{\sum_{i=1}^N (D_i - \tilde\delta ' X_i)^2 }\, ,$$
     where $\tilde{\delta}$ is the regression coefficient of $D_i$ on $X_i$. Consequently, one can approximate the distribution of $\hat \beta  -  c$ under the sharp null by considering the regression coefficients $\hat \beta_\pi$ of a linear regression in the permuted dataset $\{\frac{1}{T_1}\sum_{t \in \mathcal{T}_1}Y_{i,t} - \frac{1}{T_0}\sum_{t \in \mathcal{T}_0}Y_{i,t} - c D_{i}, D_{\pi(i)}, X_i \}_{i=1}^n$, for every permutation $\pi \in \Pi$. This amounts to running a regression on a modified dataset, where we impute the untreated potential outcome under the null, and randomly shuffle the assignment vector.

     Our procedure is also valid under weaker assumptions, in the large $N_0$-limit. To see this, let $\pi^* \sim \operatorname{Uniform}(\Pi)$, independently from the data. It is easy to see that, in the large $N_0$-limit, the regression coefficients of $D_{\pi^*(i)}$ on $X_i$ converge in probability to $0$. Consequently, we are able to show that
     
      $$\operatorname{plim}_{N_0 \to \infty} \hat \beta_{\pi^*}  -  c  =\frac{1}{N_1}\sum_{i\in (\pi^{*})^{-1}(\mathcal{J}_1)} \left(\frac{1}{T_1}\sum_{t \in \mathcal{T}_1}Y_{i,t}(0) - \frac{1}{T_0}\sum_{t \in \mathcal{T}_0}Y_{i,t}(0)\right)\, .$$

      Since, under the null, $\hat \beta  -  c$ converges in probability to $\frac{1}{N_1}\sum_{i\in \mathcal{J}_1}\left(\frac{1}{T_1}\sum_{t \in \mathcal{T}_1}Y_{i,t}(0) - \frac{1}{T_0}\sum_{t \in \mathcal{T}_0}Y_{i,t}(0)\right)$, it follows that, in the large $N_0$-limit, it suffices that the $\frac{1}{T_1}\sum_{t \in \mathcal{T}_1}Y_{j,t}(0) - \frac{1}{T_0}\sum_{t \in \mathcal{T}_0}Y_{j,t}(0)$ be iid across $j \in \mathcal{J}_0\cup \mathcal{J}_1$ for our procedure to entail asymptotically valid inference. This means that, with large $N_0$, differences in the distribution of the controls $X_i$ between treated and control arms are allowed for.

      Our procedure is also valid in the large $N_0$-limit under the same set of assumptions of \cite{conley20211inference}. This follows by noticing that, under their assumptions, the coefficient of a regression of $\frac{1}{T_1}\sum_{t \in \mathcal{T}_1}Y_{i,t} - \frac{1}{T_0}\sum_{t \in \mathcal{T}_0}Y_{i,t}$ on $X_i$ converges in probability to the coefficient $\phi$ of the conditional expectation function   $\mathbb{E}[\frac{1}{T_1}\sum_{t \in \mathcal{T}_1}Y_{j,t}(0) - \frac{1}{T_0}\sum_{t \in \mathcal{T}_0}Y_{j,t}(0)|X_i]$, which is linear in their setting.  \qed

\begin{rmk_app}
  In the spirit of Section \ref{Sec: dual}, it is possible to modify our testing procedure (both in the case with and without covariates) to account for unrestricted forms of heteroskedasticity and stochastic treatment effect heterogeneity in the large $(N_0,N_1)$ limit, while preserving validity in finite samples and in the large $N_0$-limit under the stronger assumptions of absence of both stochastic and nonstochastic treatment effect heterogeneity and homogeneity of the $\epsilon_i$, by working with studentized test statistics. See \cite{DiCiccio2017} for results in this direction,  Section 3.4.3.2 of \cite{dechaisemartin2024book} for a proposed implementation in DiD designs without covariates,  and \cite{MACKINNON2020435} for further discussion on different choices of test statistics in randomization inference in DiD settings. \qed

\end{rmk_app}

\subsection{Details on the studentized permutation $t$-test}
\label{app_perm_t}

\cite{MACKINNON2020435} suggest a permutation test as an alternative to \citeauthor{conley20211inference}'s approach to inference. In practice, their approach may be seen as a version of the permutation test discussed in Appendix \ref{app_exact}, where one relies on the test statistic:

$$\hat t = \frac{\hat \beta - c}{\sqrt{\frac{\sum_{i \in \mathcal{J}_1} ( \overline{\hat \epsilon_{i,1}} - \overline{\hat \epsilon_{i,0}})^2 }{N_1(N_1-1)} + \frac{\sum_{i \in \mathcal{J}_0} ( \overline{\hat \epsilon_{i,1}} - \overline{\hat \epsilon_{i,0}})^2 }{N_0(N_0-1)} }  }\,,$$
and the reference distribution $\{\hat t_{\pi}: \pi \in \Pi\}$, where:

$$\hat t = \frac{\hat \beta_\pi}{\hat{\sigma}_\pi}\,,$$
with
$$\hat{\sigma}_\pi = \sqrt{\frac{\sum_{i \in \mathcal{J}_1} ( \overline{\tilde \epsilon_{\pi(i),1}} - \overline{\tilde \epsilon_{\pi(i),0}})^2 }{N_1(N_1-1)} + \frac{\sum_{i \in \mathcal{J}_0} ( \overline{\tilde \epsilon_{\pi(i),1}} - \overline{\tilde \epsilon_{\pi(i),0}})^2 }{N_0(N_0-1)} }\, .  $$

Analogously to Appendix \ref{app_exact}, it follows that this test is exact for testing the sharp null \eqref{eq_sharp_ct}. However, this test is not valid for testing the null $\frac{1}{N_1 T_1} \sum_{j \in \mathcal{J}_1} \sum_{t \in \mathcal{T}_1} \tau_{j,t} = c$ under cross-sectional treatment effect heterogeneity when $N_0 \to \infty$. (In contrast, the test discussed in Appendix \ref{app_exact} is valid in the presence of deterministic heterogeneous treatment effects when $N_0\rightarrow \infty$.)

To see why the permutation $t$-test fails to control size when $N_0$ is large and there is cross-sectional heterogeneity, observe that, in this case, we have that:

$$\hat t \overset{p}{\to}    \frac{\frac{1}{N_1}\sum_{i \in \mathcal{J}_1} ( \overline{ \epsilon_{\pi(i),1}} - \overline{\epsilon_{\pi(i),0}}) + \bar{\tau}-c}{\sqrt{\frac{\sum_{i \in \mathcal{J}_1} \left( \overline{ \epsilon_{i,1}} - \overline{ \epsilon_{i,0}} + \frac{1}{T_1}\sum_{t \in \mathcal{T}_1} (\tau_{j,t}- \bar{\tau}) \right)^2}{N_1(N_1-1)}}}\, ,$$
where $\bar{\tau} = \frac{1}{N_1 T_1}\sum_{j \in \mathcal{J}_1} \sum_{t \in \mathcal{T}_1} \tau_{j,t}$.

When $N_0$ is large and $N_1$ is fixed, permutations where  residuals of treated units contribute to the denominator of $\hat{t}_{\pi}$ (i.e. permutations where $\pi(i) \in \mathcal{J}_1$ for some $i \in \mathcal{J}_1$) will constitute an asymptotically vanishing share of $|\Pi|$. As a consequence, the permutation distribution asymptotically reflects draws where  $\hat{\sigma}_\pi \approx \sqrt{\frac{\sum_{i \in \mathcal{J}_1} ( \overline{ \epsilon_{\pi(i),1}} - \overline{ \epsilon_{\pi(i),0}})^2 }{N_1(N_1-1)}} $. This will provide an incorrect approximation to the distribution of $\hat t$ under the null $\bar{\tau}=c$, unless $\frac{1}{T_1}\sum_{t \in \mathcal{T}_1} (\tau_{j,t}- \bar{\tau}) = 0$ for all $j \in \mathcal{J}_1$.

However, an advantage of this approach is that it provides asymptotically valid tests for the null $\tau^* = c$ when $N_0,N_1\to \infty$, even if one allows for heteroskedasticity, i.e. even if one relaxes the distributional parallel trends assumption \eqref{app_pt_strong} to $\mathbb{E}[\overline{{\epsilon}}_{j,1}- \overline{{\epsilon}}_{j,0}]$ being constant across $j \in \mathcal{J}_0\cup \mathcal{J}_1 $; and even if one allows for stochastic treatment effect heterogeneity. Indeed, this is the justification from \cite{MACKINNON2020435} to consider a permutation using a studentized test statistic. This result follows immediately from \cite{Janssen1997}, who shows that, in a two-sample problem, a permutation $t$-test of a sharp null of equality of distributions (which, in our setting, corresponds to the sharp null \eqref{eq_sharp_ct} under assumption \eqref{app_pt_strong}) is asymptotically valid for testing equality of means even under unequal distributions (i.e. testing $\tau^\ast = c$ under the weaker parallel trends requirement), when both samples are large. This result stands in contrast to both the original version of \cite{conley20211inference} and the implementation introduced in Appendix \ref{app_exact}, which require the strong form of parallel trends \eqref{app_pt_strong} and preclude stochastic treatment effect heterogeneity even with $N_0, N_1 \to \infty$.

\section{Appendix to Section \ref{Sec: N1>1}}
\subsection{Sign changes and the Wild Bootstrap with null imposed}
\label{app_equivalence}
\subsubsection{Relation between randomization tests under approximate symmetry and the wild bootstrap with null imposed when $N_1$ is fixed and $N_0 \to \infty$}

We consider a cross-sectional setup ($T_0=0$ and $T_1=1$) in the model-based environment of Section \ref{Sec: model based}. Specifically, we assume the researcher postulates the following model for untreated potential outcomes:

$$Y_{j}(0) = m(X_j) + \epsilon_{j} \, , \quad \mathbb{E}[\epsilon_j] = 0\, ,$$
where $m \in \mathcal{M}$ is a model for the untreated potential outcome mean. We treat the $\{X_j\}_{j=1}^N$ as fixed throughout. The researcher considers estimating the average effect on the treated by estimating a partially linear specification:

$$Y_j = \beta D_j + m(X_j) + e_j\, ,$$
and by taking as an estimator of the average effect the $\hat \beta$ that solves:

\begin{equation}
\label{eq_estimator_plm}
\hat \beta \in \operatorname{argmin}_{b \in \mathbb{R}}\frac{1}{N}\sum_{j=1}^N (Y_j - b D_j - \hat m (X_j))^2 \, ,
\end{equation}
where $\hat m$ is an estimator of $m$, e.g. the minimizer of 
\begin{equation}
\label{eq_plm_joint}
\operatorname{argmin}_{b \in \mathbb{R}, s \in \mathcal{M}_N}\frac{1}{N}\sum_{j=1}^N (Y_j - b D_j -  s (X_j))^2 \, ,
\end{equation}
with $\mathcal{M}_N$ a sieve-space that approximates $\mathcal{M}$ through a less-complex class of functions, albeit at increasing complexity with the sample size \citep{Chen2007}.

We note that the partially linear formulation is quite general. Indeed, in a DiD setting with uniform treatment timing, if one takes $Y_j$ as the pre-post difference in average outcomes, and $\mathcal{M}_N = \mathcal{M} = \mathbb{R}$, i.e. the model only accounts for an intercept, then the estimator of $\hat \beta$ in \eqref{eq_plm_joint} coincides with the two-way fixed effects estimator of the post-treatment average effect on the treated. More generally, by considering different $\mathcal{M}$, the partially linear formulation allows for more complex structures of imputation of the untreated potential outcome mean.

In this setting, the wild-bootstrap with null imposed proceeds as follows. Suppose we wish to test the null $\frac{1}{N_1} \sum_{j \in \mathcal{J}_1} \mathbb{E}[\tau_j] = c$. The researcher first estimates \eqref{eq_estimator_plm} and stores $\hat{\beta}$. She then estimates $\tilde m$ by imposing the null $\beta = c$ and use this to recover the null-imposed residuals:

$$\tilde{e}_j = Y_j - c D_j - \tilde{m}(X_j) \,, \quad i=1,\ldots, N.$$

The researcher then approximates the distribution of $\hat{\beta} - c$ under the null by considering, for each $(g_1,\ldots, g_N) \in \{-1,1\}^{N} \equiv \mathcal{G}_N$.

\begin{enumerate}
    \item Generate artificial data
    $\check Y_j^g  = c D_j +\tilde{m}(X_j) + g_j \tilde{e}_j$, $j=1,\ldots, N$.
    \item Using this artificial dataset, compute the estimator for the counterfactual mean $m$, $\check{m}^g$, and using this estimator, compute:
 $$\check{\beta}^g \in \operatorname{argmin}_{b\in \mathbb{R}}\frac{1}{N}\sum_{j=1}^n (\check{Y}_j^g - b D_j - \check{m}^g(X_j))^2$$
\end{enumerate}
The researcher then rejects the null if $\hat{\beta}$ is at the tails of $\{\check{\beta}^g - c\}_{\pi \in\mathcal{G}_n}$.

To show the asymptotic equivalence between this wild bootstrap and the randomization test discussed in the main text, we require the following assumptions.
\begin{assumption}[Consistency of $\hat m$]
\label{consistency_est}
    As $N_0 \to \infty$, the estimator $\hat m$ is consistent in the ``treatment group prediction metric'', i.e.:

    $$\max_{j \in \mathcal{J}_1} |\hat m(X_j) - m(X_j)| \overset{p}{\to} 0 \, .$$
\end{assumption}
Assumption \ref{consistency_est} requires the estimator $\hat m$ to approximate the counterfactual model for the mean in the treatment group approximately correctly (with large probability) in large samples.

The second assumption requires that information from the treated observations to not affect the estimator of $m$ with large $N_0$, in such a way that, in large samples, $\hat m$ and $\tilde m$ are approximately equivalent.

\begin{assumption}[Irrelevance of treatment group information] \label{ass_jrrelevance}
    As $N_0 \to \infty$:

    $$\max_{j \in \mathcal{J}_1} |\hat m(X_j) - \tilde m(X_j)| \overset{p}{\to} 0 \, .$$
\end{assumption}

This assumption is expected to be satisfied in a variety of settings. Observe that, for estimators of $m$ that solve \eqref{eq_plm_joint}, the contribution of the treated units to the objective function vanishes as $N_0 \to \infty$. Assumption \ref{ass_jrrelevance} requires this behaviour to be translated into the minimizer of the objective function, in such a way that the contribution of treated observations to the minimizer vanishes asymptotically.

Finally, our third assumption requires that consistent estimation of $m$ is possible with high probability across the available sign changes.

\begin{assumption}[Consistent estimation across transformations] \label{ass_cons_signs}
    Let $g^*_N \sim \operatorname{Uniform}(\mathcal{G}_N)$, independently from the data $Y_1,\ldots, Y_N$. We assume that 

    $$\lim_{N_0 \to \infty}\mathbb{P}[\max_{j \in \mathcal{J}_1} |\tilde m(X_j) - \check m^{g^*_N}(X_j)| > \delta
    ]= 0 \, , \quad \forall \delta > 0.$$
\end{assumption}

Assumption \ref{ass_cons_signs} is satisfied in a variety of settings. Suppose $m$ is estimated by  \eqref{eq_plm_joint}, with $M_N = \mathcal{M}$. We then have that $\check m^{g^*_N}$ minimizes the following objective function:

$$Q_{N}(b,s) = \frac{1}{N}\sum_{j=1}^N (\check Y_j^{g^*_N} - b D_j -  s (X_j))^2 - \frac{1}{N}\sum_{j=1}^N g_j^*\tilde{e}_{j}^2 $$

If the $\{\epsilon_{j}\}$ are iid with finite variance, $\lim_{n \to \infty} \frac{1}{N_0}\sum_{j\in \mathcal{N}_1}s(X_j)^2$ exists for every $s \in \mathcal{M}$, and $ \frac{1}{N_0}\sum_{j \in \mathcal{J}_0}|\tilde m(X_j) - m(X_j)|^2 \overset{p}{\to} 0$ (i.e. $\tilde{m}$ is consistent in a \emph{control group} average squared prediction metric), then, we have that, for every $b \in \mathbb{R}$, $s \in \mathcal{M}$, and as $N_0 \to \infty$:
\begin{equation}
    \label{eq_pointwise}
Q_{N}(b,s) = \lim_{N_0\to \infty} \frac{1}{N_0} \sum_{j \in \mathcal{J}_0}(m(X_j) - s(X_j))^2 +  o_{\mathbb{P}}(1) \, ,
\end{equation}

If $\mathcal{M} = \{c'x: c \in \mathbb{R}^{\dim(X_j)}\}$ and $\lim_{N_0 \to \infty}\frac{1}{N_0}\sum_{j\in \mathcal{J}_0} X_j X_j'$ is positive definite, then it follows from Theorem 2.7 of \cite{Newey1994} that the pointwise convergence in \eqref{eq_pointwise} is sufficient to ensure the validity of Assumption \ref{ass_cons_signs}. More generally, in nonlinear models, one should impose conditions to extend the pointwise convergence to hold uniformly over $\mathcal{M}$, which can then be combined with an M-estimator consistency result to ensure the validity of Assumption \ref{ass_cons_signs}. For sieve-estimation, Assumption \ref{ass_cons_signs} can be established by relying on consistency arguments for sieve-spaces \citep{Gallant_1987,Newey2003}. 

The following result shows the equivalence between the wild bootstrap with null imposed and the randomization test under approximate symmetry discussed in the main text.

\begin{proposition}
\label{prop_wild_bs}
    Suppose Assumption \ref{consistency_est} holds. We then have that, as $N_0 \to \infty$:
    $$\hat{\beta} \overset{p}{\to} \frac{1}{N_1}\sum_{j\in \mathcal{J}_1} (\tau_j+\epsilon_j)\, .$$

    In addition, if Assumptions \ref{ass_jrrelevance} and \ref{ass_cons_signs} also hold, then:

    $$\hat{\beta}^{g^*_N} - c \overset{p}{\to} \frac{1}{N_1}\sum_{j=1}^{N} g^*_j (\tau_j + \epsilon_j - c) \, ,$$
    where the consistency also holds conditionally at the data, i.e., for every $\delta > 0$:

    $$\mathbb{P}\left[\left|\hat{\beta}^{g^*_N} - \frac{1}{N_1}\sum_{j=1}^{N} g^*_j (\tau_j + \epsilon_j - c) \right| > \delta \Bigg|Y_1,Y_2,\ldots Y_N\right] \overset{p}{\to} 0 \, .$$
    \begin{proof}
        For the first part, note that:
        $\hat \beta  = \frac{1}{N_1}\sum_{j \in \mathcal{J}_1}(Y_j - \hat m (X_j)) = \frac{1}{N_1}\sum_{j \in \mathcal{J}_1}(\tau_j + \epsilon_j) +  \frac{1}{N_1}\sum_{j \in \mathcal{J}_1}(m(X_j) - \hat m (X_j)) = \frac{1}{N_1}\sum_{j \in \mathcal{J}_1}(\tau_j + \epsilon_j) + o_{\mathbb{P}}(1)$. For the second part, note that:
\begin{equation*}
    \begin{aligned}
        \hat{\beta}^{g^*_N}  & = \frac{1}{N_1}\sum_{j\in \mathcal{J}_1}(\check Y_j^{g^*_N} - \check m^{g^*_N}(X_j)) = c +  \frac{1}{N_1}\sum_{j\in \mathcal{J}_1} g_j^*\tilde{e}_j  +  \frac{1}{N_1}\sum_{j\in \mathcal{J}_1} (\tilde{m}(X_j) - \check m^{g^*_N}(X_j))
        \\  & = c +  \frac{1}{N_1}\sum_{j\in \mathcal{J}_1} g_j^*(\tau_j - c + \epsilon_j) + \frac{1}{N_1}\sum_{j\in \mathcal{J}_1} g_j^*(m(X_j) - \tilde{m}(X_j))   +  o_{\mathbb{P}}(1) = \\
        & =  c +  \frac{1}{N_1}\sum_{j\in \mathcal{J}_1} g_j^*(\tau_j - c + \epsilon_j) + o_{\mathbb{P}}(1) \, .
    \end{aligned}
\end{equation*}

        Finally, that the convergence also holds conditionally on the data follows from Markov inequality since, for any $\delta, l > 0$:
\begin{equation*}
    \begin{aligned}
        \mathbb{P}\left[\mathbb{P}\left[\left|\hat{\beta}^{g^*_N} - \frac{1}{N_1}\sum_{j=1}^{N} g^*_j (\tau_j + \epsilon_j - c) \right| > \delta\Bigg|Y_1,Y_2,\ldots Y_N\right]> l\right] \leq \\ 
        \frac{\mathbb{P}\left[\left|\hat{\beta}^{g^*_N} - \frac{1}{N_1}\sum_{j=1}^{N} g^*_j (\tau_j + \epsilon_j - c) \right| > \delta\right]}{l} \to 0 \, .
    \end{aligned}
\end{equation*}
    \end{proof}
\end{proposition}
\begin{rmk_app}[Clustering at a coarser level]
\label{remark_cluster}
It is apparent from the proof of Proposition \ref{prop_wild_bs} that, if one partitioned the sample into $N_1$ clusters, with one treated unit in each cluster and a common number of controls in each cluster, then, provided that Assumption \ref{ass_cons_signs} is satisfied for the cluster-level sign-changes, the asymptotic distribution of the wild bootstrap as $N_0 \to \infty$ would remain unchanged, relatively to the one obtained in  \eqref{prop_wild_bs}. This is due to the fact that, in an asymptotic framework where $N_0 \to \infty$ and $N_1$ is fixed, the contribution of controls to sampling uncertainty vanishes. Following the discussion in Supplemental Appendix S.2 of \cite{Canay_wild_bootstrap}, this suggests that, in linear specifications, clustering at a coarser level may produce a valid test with fixed $N_0$ and $N_1$ -- if the conditions in \cite{Canay_wild_bootstrap}, which include approximate symmetry, balanced clusters, and the inclusion of cluster fixed-effects in the specification, hold -- that is also valid when $N_0 \to \infty$ under the approximate symmetry conditions discussed in Section \ref{Sec: sign_changes}.
\end{rmk_app}

\begin{rmk_app}[Equivalence between clustering at a coarser level in the wild-bootstrap and clustering at a coarser level in the sign changes]
\label{remak_equiv}
In DiD specifications, when the partition of the sample into groups results in clusters with exactly the same number of treated and control units, the cluster-at-a-coarser level approach suggested in Supplemental Appendix S.2 of \cite{Canay_wild_bootstrap} is numerically identical to the partition-approach to the sign-changes test of \cite{Canay2017} discussed by the end of Section \ref{Sec: sign_changes}. This is due to the inclusion of cluster fixed effects in the specification proposed in Supplemental Appendix S.2 of \cite{Canay_wild_bootstrap}.
\end{rmk_app}

\subsubsection{``Averaging'' across partitions}
    \label{remark_avg} Still on the wild cluster bootstrap that clusters at a coarser level (Remark \ref{remark_cluster}), we note that one disadvantage of this approach is that the conclusions may be contingent on the adopted partitioning of the controls. Relatedly, tests performed with a single partitioning may entail power losses in finite samples (vis-à-vis the test that does not cluster at a coarser level), as the number of available sign changes may be substantially reduced. To mitigate these concerns, we follow an idea in \cite{song2018orderingfree} and \cite{Leung} and propose to aggregate the conclusions of the tests across diferent partitions. Suppose we want to test the null that $\tau = c$ against the two-sided alternative $\tau \neq c$. Let $\hat \beta_{g,\omega}$ denote the statistic computed under the vector of sign changes $g \in \mathcal{G}_\omega$, where $\omega$ is a choice of partition of the controls into $N_1$ equally-sized clusters, each containing one treated unit. Let $\Omega$ denote the set of all available partitions and let $\omega_1^*, \omega_1^*, \ldots \omega_S^*$ be $S$ independent  uniform draws from $\Omega$, independent from the data. If, under the null, the distribution of $\hat \beta_{g,\omega} - c$ is (approximately) invariant to the choice of $\omega \in \Omega$ and $g \in \mathcal{G}_\omega$, we may construct an aggregate decision rule as:

    $$\text{reject } H_0 \iff  \frac{1}{S}\sum_{s=1}^S \sum_{g \in \mathcal{G}_\omega}^S \frac{1}{|\mathcal{G}_\omega|} \mathbf{1}\{|\hat \beta - c| \leq |\hat \beta_{g,\omega_s^*}-c|\} \leq \alpha\, .$$

The (approximate) validity of this approach follows from (approximate) symmetry, and standard results in the theory of randomization tests (Theorem 15.2.1 of \cite{Lehmann2005} for validity under exact symmetry, and Theorem 3.1 of \cite{Canay2017} for approximate validity under approximate symmetry), since the composition between the partitioning scheme and the group of sign-changes in a given partition may be written as an (extended) group of transformations where these results apply. Importantly, by aggregating across partitions, we increase the number of available sign-changes, thus increasing power in finite samples. Observe, however, that, since the contribution of control clusters vanishes when $N_0 \to \infty$, these gains in power tend to be limited when $N_0$ is large. 

We also note that the same aggregation strategy can be adopted in the ``cluster-at-a-coarser-level'' sign-changes approach of \cite{Canay2017} discussed in the main text. However, it follows from the asymptotic equivalence result in this appendix that the aggregation strategy in both methods --- the wild cluster bootstrap with null imposed and the sign-changes test --- should perform similarly when $N_0$ is large.\footnote{In addition, it follows from Remark \ref{remak_equiv} that performance is \emph{identical} in DiD specifications, as in this case both approaches are the same.}

\subsection{\cite{Bester2011} in a DiD setting}
\label{app_bester}

\cite{Bester2011} provide conditions under which a $t$-test with cluster-robust standard errors and a degrees-of-freedom correction can be used for inference in linear models when a finite number of independent clusters is available. However, the original results in \cite{Bester2011} do not immediately apply to DiD settings where clusters are set at the treatment assignment level. Following the discussion in \cite{Canay_wild_bootstrap}, we consider the possibility of clustering at a coarser-level as a means of establishing the validity of their approach in these settings.

Specifically, we consider the following two-way fixed effects model for the untreated potential outcome $j$ in cluster $g \in \{1,2,\ldots, G\}$ at period $t$, $Y_{j,t,g}(0)$:

$$Y_{j,t,g}(0) = \lambda_j + \gamma_j + \epsilon_{j,t,g}\, .$$

We then consider the following resulting specification for observed outcomes:

\begin{equation}
\label{eq_observed}
Y_{j,t,g} = \lambda_t + \gamma_j + \tau^{\ast} D_{j,t,g} + e_{j,t,g} \,,
\end{equation}
where, as in the main text, $\tau^\ast = \frac{1}{N_1 T_1}\sum_{j\in \mathcal{J}_1}\sum_{t\in \mathcal{T}_1} \mathbb{E}[\tau_{j,t,g}]$, and $e_{j,t,g} = D_{j,t,g}(\tau_{j,t,g} - \tau^\ast) + \epsilon_{j,t,g}$. For simplicity, we assume $T_1 = T_0 = 1$. In this case, the two-way fixed-effects estimator $\hat \tau$ of $\tau^\ast$ in \eqref{eq_observed} can be obtained from the least squares estimator of the model:

$$\Delta Y_{j,2,g} = a + \tau^\ast D_{j,g} + \Delta e_{j,2,g}\, ,$$
where $D_{j,g} = \mathbf{1}\{i \in \mathcal{J}_1\}$. Let $X_{j,g}  = (1,D_{j,g})'$, $\mathcal{N}_g$ be the set of individual indices in cluster $g$, and $N_g = |\mathcal{J}_g|$ the number of individuals in cluster $g$. To apply Theorem 1 of \citet{Bester2011}, we require that $N_g = N_g'$ for every $g,g'$, that $\frac{1}{N_g} \sum_{j \in \mathcal{J}_g} X_{j,g}X_{j,g}'$ be constant across $g$, that observations in different clusters are independent, and that there exist a $2 \times 2$ covariance matrix $\Sigma$ such that (approximately):

$$\frac{1}{N_g} \sum_{j \in \mathcal{J}_g} X_{j,g} \Delta e_{j,2,g} \sim N(0, \Sigma), \quad \forall g\, .$$

If these conditions are satisfied, it follows from \cite{Bester2011} that a $t$-test based on cluster-robust standard errors (at the level of $g$) with critical values from  $\sqrt{\frac{G}{G-1}}$ times a $t_{G-1}$ random variable can be used to conduct (approximate) inference on $\tau^\ast$. 

Let us verify the above conditions in our setting. Homogeneity of $\frac{1}{N_g} \sum_{j \in \mathcal{J}_g} X_{j,g}X_{j,g}'$ is equivalent to the number of treated units being the same across clusters. Next, notice that:

$$\frac{1}{N_g} \sum_{j \in \mathcal{N}_g} X_{j,g} \Delta e_{j,2}  = \begin{pmatrix}
    \frac{1}{N_g} \sum_{j \in \mathcal{N}_g}\Delta \epsilon_{j,2,g} + p_g \frac{1}{|\mathcal{J}_1\cap \mathcal{N}_g|} \sum_{j \in \mathcal{J}_1 \cap \mathcal{N}_g}(\tau_{j,2,g} - \tau^\ast) \\
       p_g \frac{1}{|\mathcal{J}_1 \cap \mathcal{N}_g |} \sum_{j \in \mathcal{J}_1 \cap \mathcal{N}_g}(\Delta \epsilon_{j,2,g}+ (\tau_{j,2,g}-\tau^\ast))
\end{pmatrix} \, ,$$
where $p_g$ is the fraction of treated individuals in cluster $g$. Under the parallel trends assumption $\mathbb{E}[\Delta\epsilon_{j,2,g}]=0$ for every $j$, the zero-mean condition requires that

\begin{equation}
    \label{eq_rest_1}
\frac{1}{|\mathcal{J}_1\cap \mathcal{N}_g|} \sum_{j \in \mathcal{J}_1 \cap \mathcal{N}_g}(\mathbb{E}[\tau_{j,2,g}] - \tau^\ast) = 0,\quad  \forall g\,,
\end{equation}
which generally precludes deterministic heterogeneity (except in the knife-edge case where within-cluster variation in $\mathbb{E}[\tau_{j,g}] $ exactly balances out). The variance homogeneity assumption requires heteroskedasticity and treatment effect heterogeneity to be balanced across clusters. To see this, suppose within-cluster observations are independent across $j$. In this case, we have that:

\begin{equation}
\label{eq_rest_2}
    \begin{aligned}
        \mathbb{V}\left[\frac{1}{\sqrt{N_g}}\sum_{j \in \mathcal{N}_g} X_{j,g} \Delta e_{j,2}\right] = \begin{pmatrix}
    \frac{(1-p_g)^2}{|\mathcal{J}_0\cap \mathcal{N}_g|}\sum_{j \in \mathcal{J}_0\cap \mathcal{N}_g} \mathbb{V}[\Delta \epsilon_{j,2,g}] & 0 \\
    0 & 0
\end{pmatrix} +\\  \begin{pmatrix}
    p_g^2 \frac{1}{|\mathcal{J}_1\cap \mathcal{N}_g|}\sum_{j \in \mathcal{J}_1\cap \mathcal{N}_g} \mathbb{V}[\Delta \epsilon_{j,2,g} +  \tau_{j,t,g}] &   p_g^2 \frac{1}{|\mathcal{J}_1\cap \mathcal{N}_g|}\sum_{j \in \mathcal{J}_1\cap \mathcal{N}_g} \mathbb{V}[\Delta \epsilon_{j,2,g} +  \tau_{j,t,g}] \\ 
      p_g^2 \frac{1}{|\mathcal{J}_1\cap \mathcal{N}_g|}\sum_{j \in \mathcal{J}_1\cap \mathcal{N}_g} \mathbb{V}[\Delta \epsilon_{j,2,g} + \tau_{j,t,g}] &   p_g^2 \frac{1}{|\mathcal{J}_1\cap \mathcal{N}_g|}\sum_{j \in \mathcal{J}_1\cap \mathcal{N}_g} \mathbb{V}[\Delta \epsilon_{j,2,g} + \tau_{j,t,g}] \, .
\end{pmatrix}
    \end{aligned}
\end{equation}

For this matrix to be invariant across $g$, average heterogeneity should be homogeneous across clusters.

To see how the above restrictions constrain the behaviour of the within-cluster DiD estimator, note that this estimator is given by $\hat \tau_g = \frac{1}{|\mathcal{J}_1\cap \mathcal{N}_g|}\sum_{j \in \mathcal{J}_1\cap \mathcal{N}_g} \Delta Y_{j,2,g} -  \frac{1}{|\mathcal{J}_0\cap \mathcal{N}_g|}\sum_{j \in \mathcal{J}_0\cap \mathcal{N}_g} \Delta Y_{j,2,g}$. Consequently, under the parallel trends assumption $\mathbb{E}[\Delta \epsilon_{j,2,g}] = 0$ $\forall j$, requirement \eqref{eq_rest_1} is equivalent to:

$$\mathbb{E}[\hat{\tau}_g] = \tau^\ast, \quad \forall g \, .$$

Moreover, under the assumption of there being the same number of treated and control units in each cluster, the variance homogeneity requirement implies that the variance of the within-cluster DiD estimator is constant across $g$, since:

$$\hat{\tau}_g = \tau^\ast + \begin{pmatrix}-\frac{1}{1-p_g}& \frac{1}{1 - p_g}\end{pmatrix} \times \frac{1}{N_g} \sum_{j \in \mathcal{N}_g} X_{j,g}\Delta \epsilon_{j,t,g} \, .$$

Moreover,  under the assumption of there being the same number of treated and control units in each cluster, and except for knife-edge cases where variance terms exactly balance, that $\mathbb{V}[\hat{\tau}_g]$ is constant across clusters will yield that the variance of $\frac{1}{N_g} \sum_{j \in \mathcal{N}_g} X_{j,g}\Delta \epsilon_{j,t,g}$ is constant across $g$.

\begin{rmk_app}[Connection with \cite{Ibragimov2010}]
    Under the assumption that each cluster contains exactly the same number of treatment and control units, our modified version of \citeauthor{Bester2011} collapses to the proposal of \cite{Ibragimov2010}. This is due to the fact that, in this case, the two-way fixed effects estimator $\hat \tau$ is equal to the unweighted average of within-cluster DiD estimators, $\frac{1}{|\mathcal{G}|} \sum_{g \in \mathcal{G}} \hat \tau_g$; and the cluster-robust variance estimator with a degree-of-freedom correction collapses to a variance estimator that treats the within-cluster estimators as observations.
\end{rmk_app}

\begin{rmk_app}[Aggregating across available clusters]
    The application of the method of \cite{Bester2011} requires that observations be partitioned into equally-sized and balanced clusters such that the between-cluster homogeneity assumption holds. In several settings, these clusters are defined \textit{a priori}, based on known cross-unit correlation patterns. In other settings, however, there may be more than one possible cluster that satisfies the required assumptions (for example, under independently-sampled units and restrictions on heteroskedasticity). In this case, an aggregation procedure across available clusters, similar to the one in Appendix \ref{remark_avg}, may be employed. Let $\Omega$ be the set of available partitions of the units into clusters, such that the assumptions required by \cite{Bester2011} hold. Let $\hat t_{\omega,c}$ denote the $t$-test of $H_0: \tau = c$ against $H_1: \tau \neq c$ based on cluster-robust standard errors, under the partitioning scheme $\omega \in \Omega$. Denoting by $\{\omega^*_s\}_{s=1}^S$ $S$ independent and uniform draws from $\Omega$, we may construct a decision rule that aggregates across partitions as:

    $$\text{reject } H_0 \iff \frac{1}{S}\sum_{s=1}^S \mathbf{1}\left\{\hat |t_{\omega^*_s,c}| \leq \sqrt{\frac{G_s}{G_s-1}} t_{G_s-1}(1-\alpha/2)\right\}\leq \alpha\,,$$
    where $G_s$ is the number of clusters in $\omega^*_s$, and $ t_{G_s-1}(1-\alpha/2)$ is the $(1-\alpha/2)$-quantile of a t-distribution with $G_s-1$ degrees of freedom.
\end{rmk_app}

\subsection{Monte Carlo Exercise}
\label{app_mc}
We consider a simple Monte Carlo exercise in order to illustrate the properties of some of the methods discussed in the survey. We consider potential outcomes generated according to:
$$Y_{j,t}(0) = \mu_j +\lambda_t + \epsilon_{j,t}\,,$$
where the $\epsilon_{j,t}$ are iid across units, and for each $j$, generated according to a Gaussian AR(1) process with persistence $0.5$ and unit variance. We consider a DiD  estimator, and the behavior of inference methods on treatment effects under a sharp null of absence of effects, $\tau_{1,t} = 0$ for every $t \in \mathcal{T}_1$, as well as constant-treatment-effect alternatives, $\tau_{j,t} = c$ for for every $t \in \mathcal{T}_1$ and some $c > 0$. For simplicity, we fix $\mu_j = \lambda_t = 0$ for every $j$ and $t$, though  results are invariant to these values due to the use of a DiD estimator.

We contrast the behavior of four methods: a $t$-test with a cluster-robust variance estimator and the degrees-of-freedom adjustment of \cite{Ibragimov2016}; our version of the \cite{conley20211inference} approach discussed in Appendix \ref{app_exact}; the partition-and-aggregate version of the sign-changes test of \cite{Canay2017} discussed in the main text, which, following Remark \ref{remak_equiv} of  Appendix \ref{app_equivalence}, is numerically identical to the partition-and-aggregate version of the wild-cluster bootstrap in Supplemental Appendix S.2 of \cite{Canay_wild_bootstrap} that we discuss in Appendix \ref{app_equivalence}; a cluster-robust $t$-test using the jaccknife variance estimator of \cite{hansen2024jackknife} and the degrees-of-freedom adjustment of \cite{Hansen2025}. We fix $N_0=30$, $T_1=T_0=5$, and consider the behavior of tests at the 5\% significance level. We consider values $N_1 \in \{2,5,6,10\}$. Importantly, for the range of sample sizes and significance level considered, this is a setting where all methods correctly control for the type-1 error under a sharp null of no effect, even with fixed $(N_1,N_0)$.  This allows us to more transparently evaluate each approach in terms of their power under constant-treatment effect alternatives. Moreover, for this configuration of sample sizes and significance value, the approach of \cite{Ibragimov2016} is equivalent to adopting the modified critical values of \cite{potscher2023}.

\begin{figure}[H]

    \centering
\begin{subfigure}{0.45\textwidth}
      \includegraphics[scale=0.55]{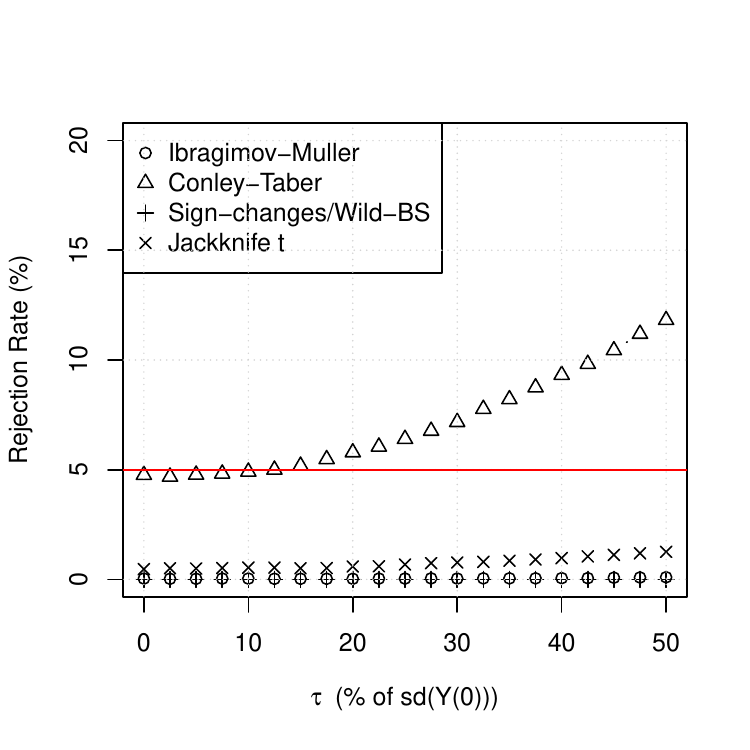}
      \caption{$N_1=2$}
     \end{subfigure}
     \begin{subfigure}{0.45\textwidth}
      \includegraphics[scale=0.55]{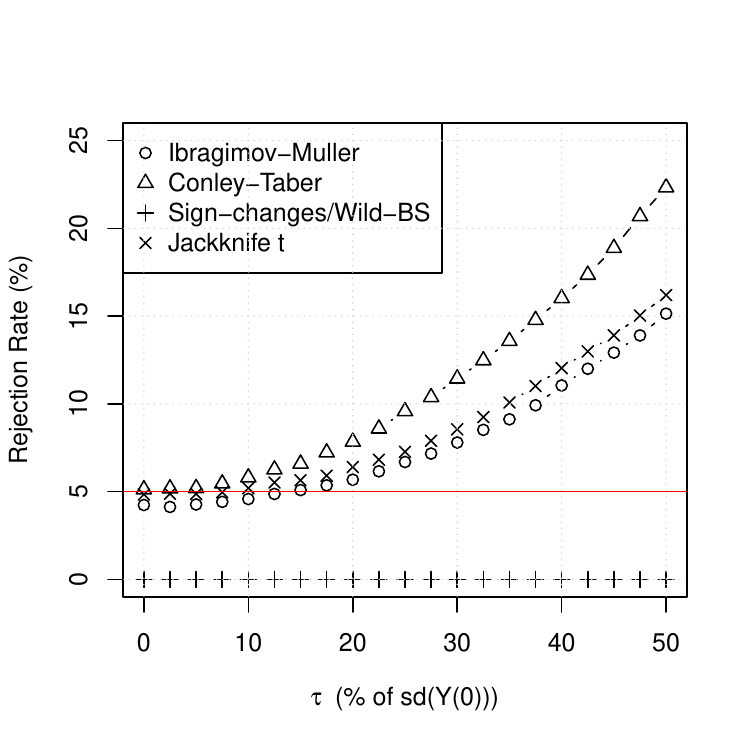}
      \caption{$N_1=5$}
     \end{subfigure}
     
          \begin{subfigure}{0.45\textwidth}
      \includegraphics[scale=0.55]{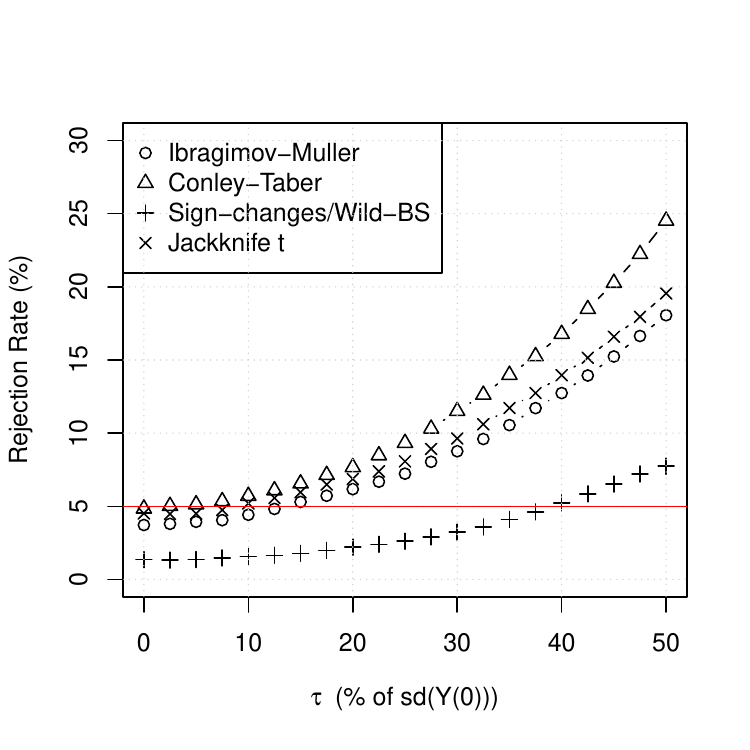}
      \caption{$N_1=6$}
     \end{subfigure}
       \begin{subfigure}{0.45\textwidth}
      \includegraphics[scale=0.55]{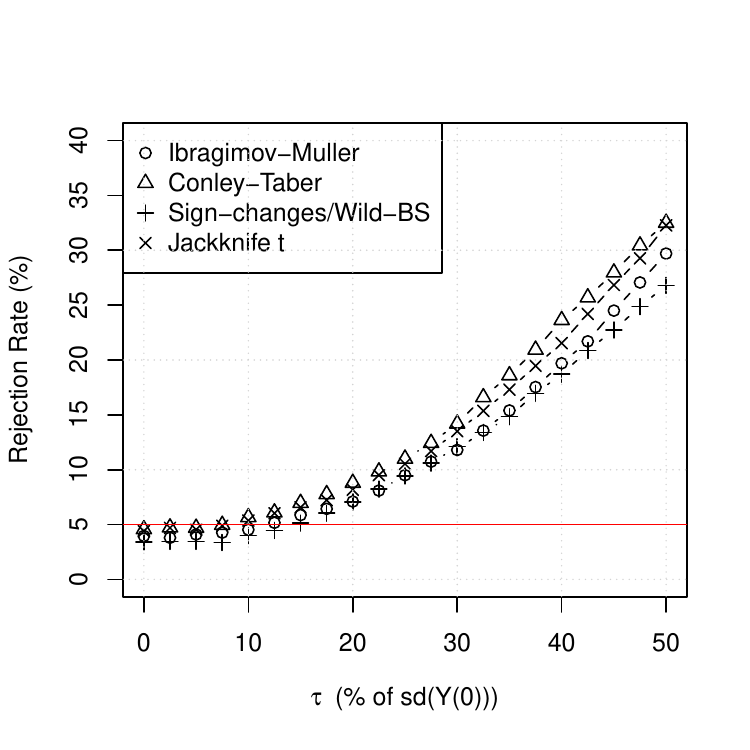}
      \caption{$N_1=10$}
     \end{subfigure}
     \caption{Rejection rates in Monte Carlo exercise}
          \label{fig_mc_results}

\end{figure}

Figure \ref{fig_mc_results} reports the results of our exercise. The red solid line in each plot indicates the nominal significance level of the tests. Each point reports the rejection rate of the corresponding method at the 5\% significance level, in a setting where treatment effects are constant and equal to a percentage of the standard deviation of $Y_{it}(0)$.

A couple of patterns are worth pointing out. First, consistent with our discussion in the main text, the sign-changes/wild-bootstrap test has trivial power when $N_1\leq 5$. Moreover, the power of the approaches that are valid when $N_1 > 1$ is always smaller than our implementation of \cite{conley20211inference}, which is valid even when $N_1=1$.\footnote{Our implementation of \cite{conley20211inference} has also some known optimality properties, being the test that, among those that control size under a sharp null of no treatment effect, has power closest to the envelope among alternatives where potential outcomes are normally distributed, and treatment effects are nonstochastic and homogeneous \citep{Lehmann1949}.}

\end{document}